\pdfoutput =0
\let \origdocument \document 
 
\def \document {\endgroup \endinput }


\title{C-RAN with Hybrid RF/FSO Fronthaul Links:\\
Joint Optimization of  RF Time Allocation and Fronthaul Compression
\vspace{-0.3cm}}
\author{Marzieh Najafi$^{\dag}$, Vahid Jamali$^{\dag}$, Derrick Wing Kwan Ng$^{\ddag}$,  and Robert Schober$^{\dag}$\\
$^{\dag}$ Friedrich-Alexander University of Erlangen-Nuremberg, Germany\\
$^{\ddag}$ University of New South Wales, Sydney, Australia
\thanks{This paper has been presented in part at IEEE Globecom 2017 \cite{MyGLOBECOM}.}
\vspace{-1.5cm}}

\maketitle

\begin{abstract}
 This paper considers the uplink of a cloud radio access network (C-RAN) comprised of several multi-antenna remote radio units (RUs) which compress the signals that they receive from multiple mobile users (MUs) and forward them to a CU via wireless fronthaul links. To enable reliable high rate fronthaul links, we employ a hybrid radio frequency (RF)/free space optical (FSO) system for fronthauling. To strike a balance between complexity and performance, we consider three different quantization schemes at the RUs, namely per-antenna vector quantization (AVQ), per-RU vector quantization (RVQ), and distributed source coding (DSC), and two different receivers at the CU, namely the linear minimum mean square error receiver and the optimal successive interference cancellation receiver.  For this network architecture, we investigate the joint optimization of the quantization noise covariance matrices at the RUs and the RF time allocation to the multiple-access and fronthaul links for rate region maximization. To this end, we formulate a unified weighted sum rate maximization problem valid for each possible pair of the considered quantization and detection schemes. To handle the non-convexity of the unified problem, we transform it into a bi-convex problem which facilitates the derivation of an efficient suboptimal solution using alternating convex optimization and golden section search.  Our simulation results show that for each pair of the considered quantization and detection schemes, C-RAN with hybrid RF/FSO fronthauling can achieve a considerable sum rate gain compared to conventional systems employing pure FSO fronthauling, especially under unfavorable atmospheric conditions. Moreover, employing a more sophisticated quantization scheme can significantly improve the system performance under adverse atmospheric conditions. In contrast, in clear weather conditions, when the FSO link capacity is high, the simple AVQ scheme performs close to the optimal DSC scheme.
\end{abstract}

\section{Introduction}
Cloud radio access network (C-RAN) is a novel cellular architecture whereby the baseband signal processing is moved from the base stations (BSs) to a cloud-computing based central unit (CU) \cite{Robert_book,C-RAN_Survey_Mao, C-RAN_Survey_VINCENTPOOR,Steve_CRAN_RoFSO}. In a C-RAN, the BSs operate as remote radio units (RUs) that relay the mobile users' (MUs') data  to the CU via fronthaul links. The CU jointly processes the MUs' data which enables the exploitation of distributed  multiple-input multiple-output (MIMO) multiplexing gains. However, conveying the signals received at the RUs to the CU via the fronthaul links is a major challenge as this may require a huge fronthaul capacity, e.g. on the order of Gbits/sec~\cite{ C-RAN_Survey_Mao}. Reviews of recent advances in fronthaul-constrained C-RANs are provided in \cite{C-RAN_Survey_Mao,C-RAN_Survey_VINCENTPOOR,Survey_C-RAN_Optical_Fronthaul}. 

In most of the existing works on C-RANs, the fronthaul links are modelled as \textit{abstract} capacity-constrained channels based on information-theoric approaches~\cite{WeiYu_JSAC, WeiYu_MIMO,WeiYu_IT}. Among practical transmission media, optical fiber has been prominently considered as a suitable candidate for fronthaul links, mainly due to the large bandwidths available at optical frequencies~\cite{C-RAN_Survey_VINCENTPOOR,Alouini_Backhaul,Survey_C-RAN_Optical_Fronthaul}. However, the implementation and maintenance of optical fiber systems are costly. Another competitive candidate technology are free space optical (FSO) systems, which provide bandwidths comparable to those of optical fiber systems, are more cost-efficient in implementation and maintenance, and are easy to upgrade~\cite{Survey_C-RAN_Optical_Fronthaul,Alouini_Backhaul,Steve_CRAN_RoFSO}. Unfortunately, the performance of FSO systems significantly deteriorates when the weather conditions are unfavorable, e.g. in snowy or foggy  weather~\cite{FSO_Survey_Murat,Survey_C-RAN_Optical_Fronthaul,Alouini_Backhaul}. On the other hand, radio frequency (RF) systems can preserve link connectivity in a more reliable manner than FSO systems but offer lower data rates. Therefore, \textit{hybrid RF/FSO} systems, where RF links are employed to support the FSO links, are appealling candidates for C-RAN fronthauling. Hybrid systems benefit from both the huge bandwidth of FSO links and the reliability of RF links~\cite{Alouini_Backhaul, MyTCOM,Survey_C-RAN_Optical_Fronthaul}. In this paper, we consider the \textit{uplink} of a C-RAN with RF multiple-access links and hybrid RF/FSO fronthaul links. As the RF spectrum is limited, we assume that the multiple-access and the fronthaul links share the same RF resources in an orthogonal manner. 

To fully exploit the fronthaul capacity, we employ quantization at the RUs. To this end, we consider three different quantization schemes, namely per-antenna vector quantization (AVQ), per-RU vector quantization (RVQ), and distributed source coding (DSC). In contrast to AVQ, RVQ  exploits the correlations between the signals received at the different antennas of a given RU, whereas DSC also takes advantage of the correlations among the signals received at different RUs \cite{Book_VQ,Book_DSC}. In addition, for the CU, we consider two different receivers, namely a linear minimum mean square error (MMSE) receiver and the optimal successive interference cancellation (SIC) receiver. The considered quantization and detection schemes offer different trade-offs between complexity and performance.  Thereby, the performance improvements of DSC over RVQ and of RVQ over AVQ come at the expense of an increased complexity. Similarly, the SIC receiver generally outperforms the linear MMSE receiver at the cost of a higher computational complexity. In practice, one can select a suitable pair of quantization and detection schemes given the affordable complexity at the RU and CU. 

The goal of this paper is to jointly optimize the quantization noise covariance matrices at the RUs and the RF time allocated to the RF multiple-access and fronthaul links for rate region maximization. In order to maximize the achievable rate region, for each pair of the adopted quantization and detection schemes, we first formulate a weighted sum rate maximization problem for optimization of the RF time allocation and the RU quantization noise covariance matrices. Then, we develop a unified representation for the resulting optimization problems valid for all considered quantization and detection schemes. Since the obtained unified optimization problem is non-convex and difficult to solve, we transform it into a bi-convex problem, i.e., a problem that is convex in each optimization variable assuming the other variables are fixed. Exploiting this property, we develop an efficient algorithm based on golden section search (GSS) and alternating convex optimization (ACO) to obtain a suboptimal solution. Moreover, we analyze the asymptotic computational complexities of the proposed algorithm and the considered quantization and detection schemes as functions of the numbers of MUs, RUs, and RU antennas.

In contrast to its conference version \cite{MyGLOBECOM}, which studies the sum rate maximization problem of an uplink C-RAN with RVQ and the optimal SIC receiver, this paper focuses on the rate region maximization problem of an uplink C-RAN employing AVQ, RVQ, or DSC at the RUs and the linear MMSE or the SIC receiver at the CU. We further note that in recent work \cite{Steve_CRAN_RoFSO}, a C-RAN architecture employing RF multiple-access links and mixed (i.e., not hybrid) RF/FSO fronthaul links was considered, where some of the RUs utilize RF fronthaul links and some use FSO fronthaul links. Moreover, the RUs apply AVQ for RF fronthauling and radio over FSO (RoFSO) for FSO fronthauling, where the RF signals are converted to the optical domain and forwarded over FSO links. The RoFSO system considered in \cite{Steve_CRAN_RoFSO} is simpler than the digital FSO system considered in this paper, however, it suffers from clipping noise, which impairs the original RF signal. 
In addition, unlike in this paper, the RF resources allocated to the multiple-access and fronthaul links in \cite{Steve_CRAN_RoFSO} are not optimized, which implies a less efficient use of the scarce RF resources.
  In the following, we briefly summarize the main contributions of this paper.

\begin{itemize}
\item We propose to employ hybrid RF/FSO systems for wireless fronthauling of C-RANs and to adaptively optimize the RF transmission time allocated to the multiple-access and fronthaul links. To the best of the authors' knowledge, this system architecture has not been considered in the literature before.

\item We investigate several quantization schemes for use at the RUs and two different detection schemes for use at the CU in order to strike a balance between complexity and performance. Moreover, we formulate a unified weighted sum rate maximization problem which is valid for any pair of the considered quantization and detection schemes and derive an efficient suboptimal solution. 
\item Our results provide several interesting insights for system design. In particular, we show that the proposed RF time allocation and fronthaul compression policies can achieve a significant performance gain compared to pure FSO fronthauling, especially under adverse weather conditions.  Furthermore, under such unfavorable conditions, applying an efficient quantization scheme at the RUs is crucial for the overall performance since the limited fronthaul capacity has to be used as effectively as possible. In contrast, under good atmospheric conditions, applying simple AVQ yields a performance close to that achieved with DSC.
\end{itemize}

The rest of this paper is organized as follows. In Section~II, we provide the system and channel models. In Section~III, we present the compression and detection strategies for the considered uplink C-RAN. In Section~IV, the rate region maximization problem is formulated and an adaptive algorithm for RF time allocation and fronthaul compression is developed. In Section~V, the complexities of the proposed algorithms and quantization and detection schemes are analyzed. Simulation results are provided in Section~VI, and conclusions are drawn in Section~VII.

\textit{ Notations:} Boldface lower-case and upper-case letters denote vectors and matrices, respectively. The superscripts $(\cdot)^{\mathsf{T}}$, $(\cdot)^{\mathsf{H}}$, and $(\cdot)^{-1}$ denote the transpose, Hermitian transpose, and matrix inverse operators, respectively; $\mathbbmss{E}\{\cdot\}$ and  $\mathrm{Tr}(\cdot)$ denote the expectation and matrix trace operators, respectively, and $|\cdot|$ denotes the determinant of a matrix  or the cardinality of a set. $\mathbf{A}\succeq \mathbf{0}$ indicates that matrix $\mathbf{A}$ is positive semidefinite;  $\mathbf{I}_n$ represents the $n$-dimentional identity matrix. Set $\mathcal{A}^{\mathsf{c}}$ denotes the complement of set $\mathcal{A}$. Moreover, $\mathbb{R}^+$, $\mathbb{R}$, and $\mathbb{C}$ denote the sets of positive real, real, and complex numbers, respectively. We use  $\mathrm{diag}\{\mathbf{A}_1,\dots,\mathbf{A}_n\}$ to denote a block diagonal matrix formed by matrices $\mathbf{A}_1,\dots,\mathbf{A}_n$. Moreover, $[x]^+$ is defined as $[x]^+\triangleq\mathrm{max}\{0,x\}$, $\ln(\cdot)$ denotes the natural logarithm, $\mathcal{O}$ is the big-O notation, and~$\mathbf{a}\sim\mathcal{CN}(\boldsymbol{\mu},\boldsymbol{\Sigma})$ is used to indicate that~$\mathbf{a}$ is a random complex Gaussian vector  with mean vector~$\boldsymbol{\mu}$ and covariance matrix~$\boldsymbol{\Sigma}$.

\section{System and Channel Models}\label{SysMod}
In this section, we present the system model and the channel models for the multiple-access and fronthaul links.\vspace{-0.3cm}

\subsection{System Model}
\begin{figure}
\centering
\scalebox{0.65}{
\pstool[width=1\linewidth]{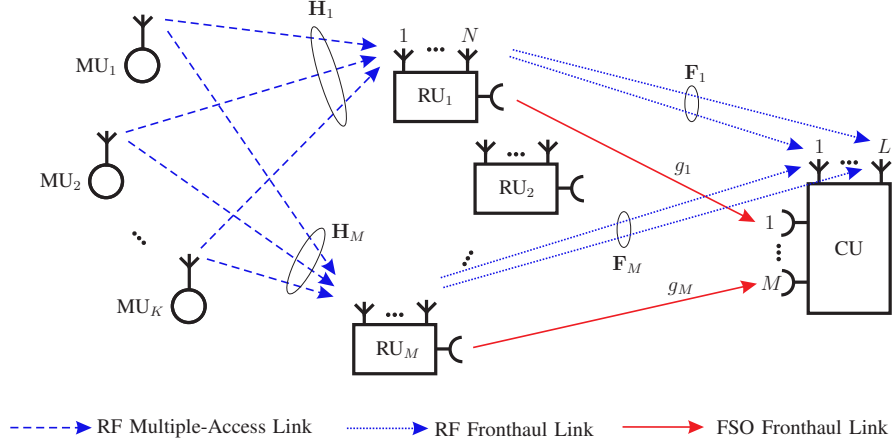}{
\psfrag{U1}[c][c][1]{MU$_1$}
\psfrag{U2}[c][c][1]{MU$_2$}
\psfrag{UK}[c][c][1]{MU$_K$}
\psfrag{H1}[c][c][1]{$\mathbf{H}_1$}
\psfrag{HM}[c][c][1]{$\mathbf{H}_M$}
\psfrag{R1}[c][c][1]{RU$_1$}
\psfrag{R2}[c][c][1]{RU$_2$}
\psfrag{RM}[c][c][1]{RU$_M$}
\psfrag{CU}[c][c][1]{CU}
\psfrag{1}[c][c][1]{$1$}
\psfrag{L}[c][c][1]{$L$}
\psfrag{M}[c][c][1]{$M$}
\psfrag{N}[c][c][1]{$N$}
\psfrag{F1}[c][c][1]{$\mathbf{F}_1$}
\psfrag{FM}[c][c][1]{$\mathbf{F}_M$}
\psfrag{g1}[c][c][1]{$g_1$}
\psfrag{gM}[c][c][1]{$g_M$}
\psfrag{Rf1}[l][c][1]{\hspace{-2mm}RF Multiple-Access Link}
\psfrag{Rf2}[l][c][1]{\hspace{-2mm}RF Fronthaul Link}
\psfrag{FSO}[l][c][1]{\hspace{-2mm}FSO Fronthaul Link}
}}
\caption{C-RAN with hybrid RF/FSO fronthaul links.}
\label{FigSysMod}\vspace{-0.3cm}
\end{figure}

We consider the uplink of a C-RAN where $K$ MUs denoted by MU$_k,\,\, k \in \mathcal{K}=\{1,\dots,K\}$, communicate with a CU via $M$ intermediate RUs denoted by RU$_m,\,\,m\in \mathcal{M}=\{1,\dots,M\}$. Fig.~\ref{FigSysMod} schematically shows the considered communication network.  We assume that the RUs and the CU are fixed nodes whereas the MUs can be mobile nodes. There is a line-of-sight between the RUs and the CU. Moreover, due to the large distance between the MUs and the CU, the direct link between them cannot support communication and is not considered. Furthermore, we assume that the RUs are half-duplex (HD) nodes with respect to (w.r.t.) the RF links. There are two transmission links in the system: $i$) the MU-RU RF multiple-access links and $ii$) the RU-CU hybrid RF/FSO fronthaul links. Each MU is equipped with a single RF antenna whereas each RU has $N $ RF antennas and an aperture FSO transmitter pointing towards the CU. The CU is equipped with $M$ photodetectors (PDs), each of which is directed to the corresponding RU, and $L$ RF antennas. We assume that the PDs are spaced sufficiently far apart such that mutual interference between the FSO links is avoided\footnote{The minimum spacing between PDs required to avoid cross talk mainly depends on the divergence angle of the FSO beams, the distance between the RUs and the CU, and the relative position of the RUs~\cite{FSO_Survey_Murat}.}. Furthermore, we assume block fading, i.e., the fading coefficients are constant during one fading block but may change from one fading block to the next.
Throughout this paper, we assume that the CU has the instantaneous channel state information (CSI) of all RF and  FSO links and is responsible for determining the transmission strategy and informing it to all nodes. Moreover, we assume that the channel states change slowly enough such that the signaling overhead caused by channel estimation and feedback is negligible compared to the amount of information transmitted in one fading block.
\subsection{Channel Model} 
 In the following, we provide the channel models for the multiple-access and fronthaul links.

\subsubsection{Multiple-Access Links}
For the RF multiple-access links, we assume a standard fading additive white Gaussian noise (AWGN) channel. All MUs transmit simultaneously using the same frequency band. The  signal received at RU$_m$ is denoted by $\mathbf{y}_m\in \mathbb{C}^{N \times 1}$ and is given by
\begin{IEEEeqnarray}{lll}\label{Eq:Signal_RF1}
\mathbf{y}_m=\mathbf{H}_m\mathbf{x}+\mathbf{n}_m,\,\,\forall m\in\mathcal{M},
\end{IEEEeqnarray}
where  $\mathbf{x}=[x_1,x_2,\dots,x_K]^{\mathsf{T}}\in \mathbb{C}^{K \times 1}$ is the vector containing the signals transmitted by all MUs and $x_k$ is the signal transmitted by MU$_k$. We assume $\mathbbmss{E}\{\mathbf{x}\mathbf{x}^{\mathsf{H}}\}=\mathrm{diag}\{P_1,P_2,\dots,P_K\}\triangleq \boldsymbol{\Sigma}$, i.e., the signals transmitted by different MUs are independent; and $P_k$ is the transmit power of MU$_k$. In addition, $\mathbf{n}_m\in \mathbb{C}^{N \times 1}$ is the noise vector at RU$_m$. The elements of $\mathbf{n}_m$, i.e., the noise samples at each antenna, are modelled as mutually independent zero-mean complex AWGN with variance $\sigma^2$. Moreover, $\mathbf{H}_m \in \mathbb{C}^{N \times K}$ denotes the channel matrix corresponding to the RF multiple-access links from the MUs to RU$_m$. 

\subsubsection{Fronthaul Links}
The fronthaul links are hybrid RF/FSO. For the FSO links, the aperture transmitter of each RU is directed towards the corresponding PD at the CU. We assume an intensity modulation direct detection (IM/DD) FSO system. 
Particularly, after removing the ambient background light intensity, the signal intensity detected at the $m$-th PD of the CU is denoted by $\tilde{{y}}_m\in \mathbb{R}$ and modelled as \cite{FSO_Vahid}
\begin{IEEEeqnarray}{lll}\label{Eq:Signal_FSO}
\tilde{y}_m={g}_m\tilde{x}_m+\tilde{n}_m,\,\,\forall m\in\mathcal{M},
\end{IEEEeqnarray}
where $\tilde{x}_m\in \mathbb{R}^+$ is the symbol transmitted by RU$_m$,  $\tilde{n}_m\in \mathbb{R}$ is the zero-mean real-valued additive white Gaussian shot noise with variance $\delta^2$  caused by ambient light at the CU, and ${g}_m\in \mathbb{R}^+$ denotes the FSO channel gain from RU$_m$ to the CU's $m$-th PD. Moreover, we assume an average power constraint $\mathbbmss{E}\{\tilde{x}_m\}\leq\tilde{P}_m$.  The capacity of the IM/DD FSO link from RU$_m$ to the CU's $m$-th PD is not known. Nevertheless, in~\cite[Eq. 26]{FSO_Cap}, the following rate has been shown to be achievable
\begin{IEEEeqnarray}{lll}\label{Eq:FSO_C}
C_m^{\mathrm{fso}}=\dfrac{1}{2}W^{\mathrm{fso}}\mathrm{log}_2\left(1+\dfrac{e\tilde{P}_m^2 g_m^2}{2\pi\delta^2}\right)\,\text{bits/sec}, \quad
\end{IEEEeqnarray}
where $W^{\mathrm{fso}}$ is the bandwidth of the FSO signal.

We assume that the multiple-access and fronthaul links utilize the same RF resources. However, since the RUs are HD nodes w.r.t. the RF links, the RF multiple-access and fronthaul links cannot be used at the same time. Therefore, we adopt a time division duplex (TDD) protocol. In particular, we divide each RF transmission time slot into $M+1$ fractions $\alpha_0,\alpha_1,\cdots,\alpha_M$ such that $\sum_{i=0}^M\alpha_{i}=1$ holds, where in $\alpha_0$ fraction of each time slot, the RF multiple-access links are active and in $\alpha_m,\forall m\in\mathcal{M}$, fraction, the RF fronthaul link from RU$_m$ to the CU is active. The RF signal of RU$_m$ received at the CU is denoted by $\bar{\mathbf{y}}_m\in \mathbb{C}^{L \times 1}$ and given by
\begin{IEEEeqnarray}{lll}\label{Eq:Signal_RF2}
\bar{\mathbf{y}}_m=\mathbf{F}_m\bar{\mathbf{x}}_m+\bar{\mathbf{n}}_m,\,\,\forall m\in\mathcal{M},
\end{IEEEeqnarray}
where $\bar{\mathbf{x}}_m\in \mathbb{C}^{N \times 1}$ is the vector of signals transmitted over the antennas of RU$_m$. We assume that $\mathbbmss{E}\{\bar{\mathbf{x}}^{\mathsf{H}}_m\bar{\mathbf{x}}_m\}=\bar{P}_m$ holds, where $\bar{P}_m$ is the fixed transmit power of RU$_m$ over the RF fronthaul link. $\bar{\mathbf{n}}_m\in \mathbb{C}^{L \times 1}$ denotes the noise vector at the CU. The noise at each antenna of the CU, i.e., each element of $\bar{\mathbf{n}}_m$, is modelled as zero-mean complex AWGN with variance $\varrho^2$. Moreover, $\mathbf{F}_m\in \mathbb{C}^{L \times N}$ denotes the channel matrix of the RF fronthaul link from RU$_m$ to the CU. The capacity of the RF fronthaul link between RU$_m$ and the CU is obtained via optimal waterfilling as \cite{FSO_Vahid}
\begin{IEEEeqnarray}{lll}\label{Eq:RF_C}
C_{m}^{\mathrm{rf}}=W^{\mathrm{rf}}\sum_{j=1}^{\min\{N,L\}}\left[\mathrm{log_2}\left\{\dfrac{\mu\chi_{m,j}^2}{\varrho^2}\right\}\right]^+\,\text{bits/sec}, \quad
\end{IEEEeqnarray}
 where  $W^{\mathrm{rf}}$ is the bandwidth of the  RF signal. In (\ref{Eq:RF_C}), $\chi_{m,j}$ is the $j$-th singular value of $\mathbf{F}_m$ and $\mu$ is the water level which is chosen to satisfy the power constraint as the solution of the following equation
\begin{IEEEeqnarray}{lll}\label{Eq:mu}
\sum_{j=1}^{\min\{N,L\}}\left[\mu-\dfrac{\varrho^2}{\chi_{m,j}^2}\right]^+=\bar{P}_m.
\end{IEEEeqnarray}

\section{Compression, Transmission, and Detection Strategies}

\begin{figure}
\centering
\scalebox{0.6}{
\pstool[width=0.9\linewidth]{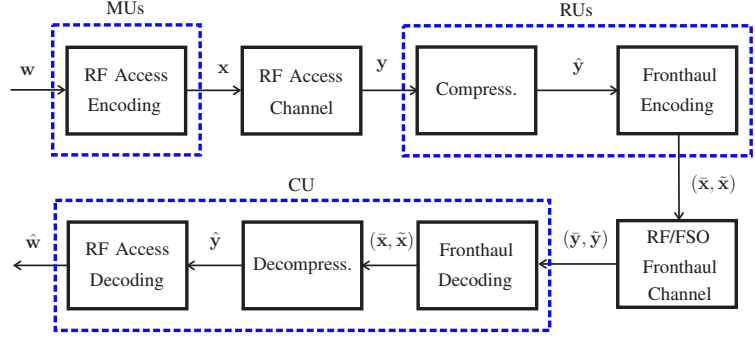}{
\psfrag{w}[c][c][1]{$\mathbf{w}$}
\psfrag{hw}[c][c][1]{$\hat{\mathbf{w}}$}
\psfrag{x}[c][c][1]{$\mathbf{x}$}
\psfrag{Y}[c][c][1]{$\mathbf{y}$}
\psfrag{hY}[c][c][1]{$\hat{\mathbf{y}}$}
\psfrag{xx}[c][c][1]{$(\bar{\mathbf{x}},\tilde{\mathbf{x}})$}
\psfrag{yy}[c][c][1]{$(\bar{\mathbf{y}},\tilde{\mathbf{y}})$}
\psfrag{MU}[c][c][1]{MUs}
\psfrag{RU}[c][c][1]{RUs}
\psfrag{CU}[c][c][1]{CU}
\psfrag{RF}[c][c][1]{RF Access}
\psfrag{CH}[c][c][1]{Channel}
\psfrag{Cmp}[c][c][1]{Compress.}
\psfrag{FR}[c][c][1]{Fronthaul}
\psfrag{En}[c][c][1]{Encoding}
\psfrag{HR}[c][c][1]{Hybrid}
\psfrag{FS}[c][c][1]{RF/FSO}
\psfrag{FrDe}[c][c][1]{Fronthaul Decoding}
\psfrag{Decom}[c][c][1]{Decompress.}
\psfrag{De}[c][c][1]{Decoding}
}}\vspace{-0.3cm}
\caption{Block diagram of the considered uplink C-RAN system. $\mathbf{w}$ is the vector of the messages of the MUs and $\hat{\mathbf{w}}$ is the corresponding estimate at the CU. Moreover, $\mathbf{\tilde{x}}=[\tilde{x}_1,\tilde{x}_2,\dots,\tilde{x}_M]^{\mathsf{T}}$ and $\mathbf{\bar{x}}=[\mathbf{\bar{x}}_1^{\mathsf{T}},\mathbf{\bar{x}}_2^{\mathsf{T}},\dots,\mathbf{\bar{x}}_M^{\mathsf{T}}]^{\mathsf{T}}$ denote the vectors containing the symbols transmitted by the RUs over the FSO links and the RF fronthaul links, respectively.}
\label{FigBlockD}\vspace{-0.3cm}
\end{figure}

In this section, we first present the three considered compression schemes for the RUs, then we introduce the transmission strategy over the fronthaul links, and finally, we explain the two considered detection schemes for the CU. The block diagram of the considered uplink C-RAN is depicted in Fig.~\ref{FigBlockD}.

\subsection{Fronthaul Compression}
We assume that the RUs employ compress-and-forward relaying and quantize their received signals. Let ${\mathbf{y}}=[{\mathbf{y}}_1^{\mathsf{T}},\dots,{\mathbf{y}}_M^{\mathsf{T}}]^{\mathsf{T}}$ and $\hat{\mathbf{y}}=[\hat{\mathbf{y}}_1^{\mathsf{T}},\dots,\hat{\mathbf{y}}_M^{\mathsf{T}}]^{\mathsf{T}}$ denote the vectors of received and quantized signals at all RUs where $\hat{\mathbf{y}}_m$ is the quantized version of RU$_m$'s signal. In particular, the RUs quantize their received imphase/quadrature (I/Q) samples with a sampling rate of $f_s\geq W^{\mathrm{rf}} $ and forward the compressed signals to the CU through hybrid RF/FSO fronthaul links. Assuming the Gaussian quantization test channel, $\hat{\mathbf{y}}$ is given by~\cite{WeiYu_JSAC}\vspace{-0.5cm}
\begin{IEEEeqnarray}{lll}\label{Eq:noise_Q_global}
\hat{\mathbf{y}}=\mathbf{y}+\mathbf{z},
\end{IEEEeqnarray}
where $\mathbf{z}\in \mathbb{C}^{MN\times 1}\sim \mathcal{CN}(\mathbf{0}_{MN}, \mathbf{D})$ is the quantization noise and $\mathbf{D}=\mathbbmss{E}\{\mathbf{z}\mathbf{z}^{\mathsf{H}}\}$ is the joint distortion matrix at the RUs. The mean square distortion between the received I/Q sample vector $\mathbf{y}$  and the corresponding quantized vector $\hat{\mathbf{y}}$ is given by the main diagonal entries of $\mathbf{D}$. In this paper, we consider three commonly used quantization schemes, namely AVQ, RVQ, and DSC, which offer different trade-offs between complexity and performance.

\subsubsection{Per-Antenna Vector Quantization}
One simple way of compressing the received signals at the RUs is to independently quantize the I/Q samples at each antenna of the RUs. In this case, $\mathbf{D}$ is a diagonal matrix. Let $d_{m,n}$ be the distortion at the $n$-th antenna of RU$_m$, i.e., $d_{m,n}=\mathbbmss{E}\{|\hat{y}_{m,n}-y_{m,n}|^2\}$, where  $\hat{y}_{m,n}$ and $y_{m,n}$ denote the quantized and the original versions of the signal received at the $n$-th antenna of RU$_m$, respectively. Therefore, $\mathbf{D}$ is obtained as $\mathbf{D}=\mathrm{diag}\{d_{1,1},\dots,d_{1,N},\dots,d_{M,1},\dots,d_{M,N}\}$. 
Moreover, based on the lossy source coding theorem~\cite{ElGamal}, the output rate of the quantizers at RU$_m$, denoted by $r_m$, has to meet the following constraint in order to guarantee a maximum distortion of $d_{m,n}$
\begin{IEEEeqnarray}{lll}\label{Eq:SQ_SC}
r_m\geq \alpha_0 f_s\sum_{\forall n} I\left(  y_{m,n};\hat{y}_{m,n}\right),\,\,\forall m\in\mathcal{M},
\end{IEEEeqnarray}
where $I\left(  y_{m,n};\hat{y}_{m,n}\right)$ is given by
\begin{IEEEeqnarray}{lll}\label{Eq:MI_SQ}
I\left(  y_{m,n};\hat{y}_{m,n}\right)=\mathrm{log_2}\dfrac{ \mathbf{h}_{m,n}^{\mathsf{T}}\boldsymbol{\Sigma}\mathbf{h}_{m,n}^{\mathsf{*}}+d_{m,n}+\sigma^2}{d_{m,n}}.
\end{IEEEeqnarray}
Here, $\mathbf{h}_{m,n}\in \mathbb{C}^{K\times 1}$ is a vector containing the channel coefficients between the MUs and the $n$-th antenna of RU$_m$.

\subsubsection{Per-RU Vector Quantization}
A higher compression rate can be achieved if each RU exploits the correlation between the signals it receives at different antennas and jointly quantizes the I/Q samples of the antennas. In this case, $\mathbf{D}$ is a block diagonal matrix with matrix entries $\mathbf{D}_m,\,\forall m\in\mathcal{M}$, where $\mathbf{D}_m$, i.e., the covariance matrix of $\mathbf{z}_m=\hat{\mathbf{y}}_m-\mathbf{y}_m$ at RU$_m$, is in general non-diagonal. Moreover, according to the lossy source coding theorem~\cite{ElGamal}, the output rate of the quantizer at RU$_m$ has to satisfy the following constraint if the original signal $\mathbf{y}_m$ is to be recovered with maximum distortion $\mathbf{D}_m$ 
\begin{IEEEeqnarray}{lll}\label{Eq:VQ_SC}
r_m\geq \alpha_0 f_sI\left(\mathbf{y}_m;\hat{\mathbf{y}}_m\right),\,\,\forall m\in\mathcal{M},
\end{IEEEeqnarray}
where $I\left(\mathbf{y}_m;\hat{\mathbf{y}}_m\right)$ is given by
\begin{IEEEeqnarray}{lll}\label{Eq:MI_VQ}
I\left(\mathbf{y}_m;\hat{\mathbf{y}}_m\right) = \mathrm{log_2}\dfrac{\left\vert \mathbf{H}_m\boldsymbol{\Sigma}\mathbf{H}_m^{\mathsf{H}}+\mathbf{D}_m+\sigma^2\mathbf{I}_{N}\right\vert}{\left\vert \mathbf{D}_m\right\vert}.
\end{IEEEeqnarray}

\subsubsection{Distributed Source Coding}
Since the received signals at different RUs are statistically dependent, the RUs can exploit this dependence via DSC. It has been shown that multiple isolated sources, i.e., the RUs here, can compress data using DSC as efficiently as if they knew each others'  received signals and jointly quantized them \cite{ElGamal,Cover,DistributedSourceCoding}. In DSC, each RU quantizes its received signal based on the conditional statistics of (or the correlation model between) the RUs' signals and not based on the actual signals received at the other RUs~\cite{DistributedSourceCoding}. Let $\mathbf{D}$ denote the joint distortion matrix of the signals received at all RUs. Based on the lossy source coding theorem~\cite{ElGamal}, for DSC, the output rates of the quantizers at the RUs have to meet the following constraints
\begin{IEEEeqnarray}{lll}\label{Eq:DSC_SC}
\sum_{\forall m\in\mathcal{S}}r_m\geq \alpha_0 f_sI\left(  \mathbf{y}(\mathcal{S});\hat{\mathbf{y}}(\mathcal{S})|\hat{\mathbf{y}}(\mathcal{S}^{\mathsf{c}})\right),\,\,\forall \mathcal{S}\subseteq\mathcal{M},
\end{IEEEeqnarray}
where in vectors $\mathbf{y}(\mathcal{S})$ and $\hat{\mathbf{y}}(\mathcal{S})$, we collect all $\mathbf{y}_m$ and $\hat{\mathbf{y}}_m$ for $m\in\mathcal{S}$, respectively, and $ \mathcal{S}$ is a non-empty subset of $\mathcal{M}$. Moreover, $I\left(  \mathbf{y}(\mathcal{S});\hat{\mathbf{y}}(\mathcal{S})\mid\hat{\mathbf{y}}(\mathcal{S}^{\mathsf{c}})\right)$ is given by
\begin{IEEEeqnarray}{lll}\label{Eq:MI_DSC}
I\left(  \mathbf{y}(\mathcal{S});\hat{\mathbf{y}}(\mathcal{S})\mid\hat{\mathbf{y}}(\mathcal{S}^{\mathsf{c}})\right)  
 =  \mathrm{log_2}\dfrac{\left\vert \mathbf{H}(\mathcal{S})\boldsymbol{\Sigma}(\mathbf{H}(\mathcal{S}))^{\mathsf{H}}+\mathbf{D}(\mathcal{S})+\sigma^2\mathbf{I}_{|\mathcal{S}|N}\right\vert}{\left\vert \mathbf{D}(\mathcal{S})\right\vert},
\end{IEEEeqnarray}
where matrices $\mathbf{H}(\mathcal{S})$ and $\mathbf{D}(\mathcal{S})$ contain all $\mathbf{H}_m$ and $\mathbf{D}_m$, respectively, whose indices $m$ belong to set $\mathcal{S}$.

\subsection{Fronthaul Transmission}
Due to the limited available RF spectrum, we assume that both the multiple-access and the fronthaul RF links utilize the same RF resource. Hence, the advantages of employing a hybrid RF/FSO system for fronthauling come at the expense of a bandwidth reduction for the multiple-access links. In the following, we propose an adaptive protocol which divides the available RF transmission time between the multiple-access and the fronthaul links in a TDD manner\footnote{Although TDD based time sharing of RF resources between multiple-access and fronthaul links is in general suboptimal, we adopt this approach here to avoid interlink interference, which facilitates a simple transceiver design.}. 
 Recall that $\alpha_m \in [0,1]$ fraction of each RF time slot is allocated to the fronthaul link from RU$_m$ to the CU. Hence, a fraction of $\alpha_0 = 1-\sum_{\forall m}\alpha_{m}$ of each RF time slot is available for the multiple-access links. For future reference, we define $\boldsymbol{\alpha}=[\alpha_0,\alpha_1,\dots,\alpha_M]^\mathsf{T}\in\mathcal{A}$, where $\mathcal{A}=\big\{\boldsymbol{\alpha}\mid\sum_{m=0}^{M}\alpha_m=1,\,\alpha_m\in[0,1],\,\, \forall m \in \{0,1,\dots,M\}\big\}$.
In order to recover $\hat{\mathbf{y}}_m$ reliably at the CU, based on the channel coding theorem~\cite{ElGamal,Cover}, the rate at the output of the quantizers at RU$_m$ has to meet the following constraint
\begin{IEEEeqnarray}{lll}\label{Eq:CC}
r_m\leq  {C}_m^{\mathrm{fso}}+\alpha_m{C}_m^{\mathrm{rf}},\,\,\forall m \in \mathcal{M},
\end{IEEEeqnarray}
where ${C}_m^{\mathrm{fso}}$ and ${C}_m^{\mathrm{rf}}$ are given in (\ref{Eq:FSO_C}) and (\ref{Eq:RF_C}), respectively.

\subsection{Detection at the CU}
For detection of the signals transmitted by the MUs at the CU, we consider two detection schemes, namely a suboptimal linear detector and an optimal detector. The linear detector requires lower computational complexity compared to the optimal detector at the expense of a loss in performance. In the following, we explain the detectors employed at the CU in detail. For future reference, let $\mathbf{h}_{k}\in \mathbb{C}^{ MN\times 1}$ denote the channel vector corresponding to the RF multiple-access links from MU$_k$ to all RUs. 
 
\subsubsection{Linear Detector}
We assume that $\mathbf{m}_{k}\in \mathbb{C}^{ MN\times 1}$ is a linear filter applied at the CU in order to recover MU$_k$'s signal $x_k$. The corresponding output signal of the filter is given by $\mathbf{m}_{k}^{\mathsf{H}} \hat{\mathbf{y}}$. Let $\gamma_k$ denote the signal-to-interference-plus-noise ratio (SINR) at the output of the  filter for MU$_k$ which is given by
\begin{IEEEeqnarray}{lll}\label{Eq:SINR_k}
\gamma_k=\dfrac{P_k\mathbf{m}_{k}^{\mathsf{H}} \mathbf{h}_{k}\mathbf{h}_{k}^{\mathsf{H}}\mathbf{m}_{k}}{\mathbf{m}_{k}^{\mathsf{H}}\left( \sum_{j\neq k}^{K}P_j\mathbf{h}_{j}\mathbf{h}_{j}^{\mathsf{H}}+\mathbf{D}+\sigma^2\mathbf{I}_{MN}\right)\mathbf{m}_{k}}.
\end{IEEEeqnarray}
The linear filter that maximizes $\gamma_k$, denoted by $\mathbf{m}_{k}^{\mathrm{sinr}}$, and the maximum SINR $\gamma_k^{\mathrm{opt}}$ can be obtained using the Rayleigh quotient inequality~\cite{Rayleigh_Quotient} as
\begin{IEEEeqnarray}{lll}\label{Eq:MMSE Filter}
\mathbf{m}_{k}^{\mathrm{sinr}}&=\left(\sum_{j\neq k}^{K}P_j\mathbf{h}_{j}\mathbf{h}_{j}^{\mathsf{H}}+\mathbf{D}+\sigma^2\mathbf{I}_{MN}\right)^{-1}\mathbf{h}_k\quad \text{and}\\
\gamma_k^{\mathrm{opt}} &= P_k\mathbf{h}_{k}^{\mathsf{H}}\left(\sum_{j\neq k}^{K}P_j\mathbf{h}_{j}\mathbf{h}_{j}^{\mathsf{H}}+\mathbf{D}+\sigma^2\mathbf{I}_{MN}\right)^{-1}\mathbf{h}_{k},\label{Eq:MMSE SINR}
\end{IEEEeqnarray}
respectively. As can be observed, (\ref{Eq:MMSE SINR}) is identical to the SINR at the output of an MMSE detector~\cite{MMSE}. Hence, the optimal linear detector that maximizes the MUs' SINRs is the well-known linear MMSE detector. Therefore, for the Gaussian RF access channel in (\ref{Eq:Signal_RF1}) and the Gaussian quantization test channel in (\ref{Eq:noise_Q_global}), the transmission rate of MU$_k$ employing linear MMSE detection at the CU is obtained as \cite{WeiYu_JSAC}
\begin{IEEEeqnarray}{lll}\label{Eq:R_k_MMSE1}
R_{k}\left(\alpha_0,\mathbf{D}\right)=\alpha_0 W^{\mathrm{rf}}  I\left({x}_k;(\mathbf{m}_{k}^{\mathrm{sinr}})^{\mathsf{H}}\hat{\mathbf{y}}\right)= \alpha_0 W^{\mathrm{rf}} \mathrm{log_2}\left(1+\gamma_k^{\mathrm{opt}}\right).\quad
\end{IEEEeqnarray}
Using Sylvester's determinant theorem~\cite{Horn_Matrix}, (\ref{Eq:R_k_MMSE1}) can be written as follows
\begin{IEEEeqnarray}{lll}\label{Eq:R_k_MMSE}
R_{k}\left(\alpha_0,\mathbf{D}\right)= \alpha_0 W^{\mathrm{rf}} \mathrm{log_2} \dfrac{\left\vert\sum_{j=1}^{K}P_j\mathbf{h}_{j}\mathbf{h}_{j}^{\mathsf{H}}+\mathbf{D}+\sigma^2\mathbf{I}_{MN}\right\vert}{\left\vert \sum_{j\neq k}^{K}P_j\mathbf{h}_{j}\mathbf{h}_{j}^{\mathsf{H}}+\mathbf{D}+\sigma^2\mathbf{I}_{MN}\right\vert}.
\end{IEEEeqnarray}

\subsubsection{Optimal Detector}
For the multiple access channel (MAC), SIC is capacity-achieving. Assume an arbitrary decoding order of $1,\dots,K$, and suppose the MUs are indexed according to the decoding order. The SIC decoder first decodes the signal of MU$_1$, subtracts it from the combined signal, $\hat{\mathbf{y}}$, and continues until all MUs' signals $x_1,\dots,x_K$ are decoded. Therefore, the signal based on which the message of MU$_k$ is decoded is given by $\hat{\mathbf{y}}^{\mathrm{sic}}_k=\hat{\mathbf{y}}-\sum_{i=1}^{k-1}x_i\mathbf{h}_i$. For the Gaussian RF MAC in (\ref{Eq:Signal_RF1}) and the Gaussian quantization test channel in (\ref{Eq:noise_Q_global}), the transmission rate of MU$_k$ employing SIC at the CU is obtained as~\cite{WeiYu_JSAC}

\begin{IEEEeqnarray}{lll}\label{Eq:R_k_S}
R_{k}\left(\alpha_0,\mathbf{D}\right)=\alpha_0 W^{\mathrm{rf}}  I\left({x}_k;\hat{\mathbf{y}}^{\mathrm{sic}}_k\right)= \alpha_0 W^{\mathrm{rf}} \mathrm{log_2}\dfrac{\left\vert \sum_{j=k}^{K}P_j\mathbf{h}_{j}\mathbf{h}_{j}^{\mathsf{H}}+\mathbf{D}+\sigma^2\mathbf{I}_{MN}\right\vert}{\left\vert \sum_{j> k}^{K}P_j\mathbf{h}_{j}\mathbf{h}_{j}^{\mathsf{H}}+\mathbf{D}+\sigma^2\mathbf{I}_{MN}\right\vert}.
\end{IEEEeqnarray}

\begin{remk}
Employing SIC to achieve (\ref{Eq:R_k_S}) is equivalent to applying MMSE decision-feedback equalization to the signal received at the CU. This is due to the fact that for Gaussian inputs and the AWGN channel, the MMSE filter is information lossless~\cite[Proposition~5.13]{MMSE_lossless}. In other words, $I(x_k;\hat{\mathbf{y}}^{\mathrm{sic}}_k)=I(x_k;(\mathbf{m}_{k}^{\mathrm{sic}})^{\mathsf{H}}\hat{\mathbf{y}}^{\mathrm{sic}}_k)$ holds, where $\mathbf{m}_{k}^{\mathrm{sic}}=\left(\sum_{j>k}^{K}P_j\mathbf{h}_{j}\mathbf{h}_{j}^{\mathsf{H}}+\mathbf{D}+\sigma^2\mathbf{I}_{MN}\right)^{-1}\mathbf{h}_k$ is the MMSE filter after removing MU$_j$'s signal $\forall j<k$. 
\end{remk}

\section{Rate Region Maximization}
In this section, we first formulate the rate region maximization problems for the considered quantization and detection schemes, and then represent them in a unified form. Since the resulting unified problem is non-convex, we transform the problem to a bi-convex problem which enables the derivation of an effective suboptimal solution. Based on this solution, we propose an adaptive protocol for joint optimization of the fronthaul compression and the RF time allocation.

\subsection{Problem Formulation}
Let $\mathbf{r}=[r_1,\dots,r_M]^{\mathsf{T}}$ be the vector of the output rates of the quantizers at the RUs. Moreover, let $\boldsymbol{\mu}=[\mu_1,\dots,\mu_K]^{\mathsf{T}}$, where $\mu_{k}$ is the weight representing the priority associated with MU$_k$. Without loss of generality, we assume that $\mu_k\geq0,\,\forall k\in\mathcal{K}$, and $\sum_{k=1}^{K}\mu_k=1$. Then, the rate region maximization problem can be formulated as the following weighted sum rate maximization problem
\begin{IEEEeqnarray}{lll}\label{Eq:RateRegion}
&\underset{\boldsymbol{\alpha}\in \mathcal{A},\mathbf{D}\succeq \mathbf{0},\mathbf{r}}{\mathrm{maximize}}\,\,&\sum_{k=1}^{K} \mu_{k} R_k\left(\alpha_0,\mathbf{D}\right) \\
 &\mathrm{subject\,\, to\,\,} &\mathrm{C}1:\, \mathbf{r}\,\,\text{satisfies}\,\,(\ref{Eq:SQ_SC})\,\,  \text{or} \,\, (\ref{Eq:VQ_SC})\,\, \text{or}\, (\ref{Eq:DSC_SC}),\nonumber\\
 &&\mathrm{C}2:\,\mathbf{r}\,\,\text{satisfies}\,\,(\ref{Eq:CC}),\nonumber
  \end{IEEEeqnarray}
where $R_k\left(\alpha_0,\mathbf{D}\right)$ is the transmission rate of MU$_k$ given in (\ref{Eq:R_k_MMSE}) and (\ref{Eq:R_k_S}) for the MMSE and SIC receivers at the CU, respectively. Moreover, constraints (\ref{Eq:SQ_SC}), (\ref{Eq:VQ_SC}), and  (\ref{Eq:DSC_SC}) in $\mathrm{C}1$ are the source coding constraints for AVQ, RVQ, and DSC, respectively, and (\ref{Eq:CC}) in $\mathrm{C}2$ represents the channel capacity constraint.

To fully characterize the rate region, we have to solve (\ref{Eq:RateRegion}) for all possible $\boldsymbol{\mu}$. Note that for a given $\boldsymbol{\mu}$, the optimal SIC order is obtained by decoding the MUs with smaller weights first~\cite{WeiYu_MIMO}. Thereby, without loss of generality, we assume $0\leq\mu_{1}\leq \dots\leq \mu_{K}$ resulting in the decoding order  $1,\dots,K$ in the following.

Next, we provide a unified representation of (\ref{Eq:RateRegion}) which enables us to solve this problem for all considered quantization and detection schemes efficiently. To this end, we first define some auxiliary variables. Let $\mathcal{T}$ denote a subset of $\mathcal{N}_T=\{1,2,\dots,MN\}$, where $\mathcal{N}_T$ is the index set of all antennas at the RUs. Moreover, we define $\mathcal{T}_{\mathcal{S}}=\{(m-1)N+1,\dots,mN|\forall m\in \mathcal{S}\}$ as the index set corresponding to all antennas of RU$_m,\forall m\in\mathcal{S}$, where $\mathcal{S}$ is a non-empty subset of $\mathcal{M}$. Furthermore, we define $\mathbf{I}(\mathcal{A})$ for any given set $\mathcal{A}$ as the pseudo identity matrix of size $|\mathcal{A}| \times MN$ whose entry in the $i$-th row and $j$-th column is equal to $1$ if the $i$-th element of set $\mathcal{A}$ is equal to $j$, otherwise it is equal to zero. 
Based on the definition of $\mathbf{I}(\mathcal{A})$, we use $\mathbf{I}(\mathcal{A})\mathbf{D}(\mathbf{I}(\mathcal{B}))^{\mathsf{T}}$ to select the $(i,j)$-th entries of $\mathbf{D},$  $\forall i\in\mathcal{A}$ and $\forall j\in\mathcal{B}$. For instance, for $\mathcal{A}=\{1,2\}$ and $\mathcal{B}=\{1,3\}$, $\mathbf{I}(\mathcal{A})\mathbf{D}(\mathbf{I}(\mathcal{B}))^{\mathsf{T}}=\begin{bmatrix}
d_{11} &d_{13}\\
d_{21} &d_{23}
\end{bmatrix} $ holds.

Using the above notations, in the following proposition,  we present (\ref{Eq:RateRegion}) for all adopted quantization and detection schemes in a unified manner.

\begin{prop}\label{Unify}
The achievable rate region maximization problem (\ref{Eq:RateRegion}) can be formulated in unified form for the considered quantization and detection schemes as follows
\begin{IEEEeqnarray}{llrl}\label{Eq:Unified}
 &\underset{\boldsymbol{\alpha}\in \mathcal{A},\mathbf{D},\mathbf{r}}{\mathrm{maximize}}&&\alpha_0 W^{\mathrm{rf}}\sum_{k=1}^{K} \mu_{k}\mathrm{log_2}\dfrac{\left\vert\mathbf{V}_k+\mathbf{D}\right\vert}{\left\vert \mathbf{W}_k+\mathbf{D}\right\vert}\nonumber \\
& \mathrm{subject\,\, to} \,\, &{\mathrm{C}1:} \,\,\, & \sum_{m\in\mathcal{S}}r_m \geq \alpha_0 f_s \mathrm{log_2}\dfrac{\left\vert \mathbf{C}(\mathcal{S})+\mathbf{I}(\mathcal{T}_{\mathcal{S}})\mathbf{D}(\mathbf{I}(\mathcal{T}_{\mathcal{S}}))^{\mathsf{T}}+\sigma^2\mathbf{I}_{|\mathcal{S}|N}\right\vert}{\left\vert \mathbf{I}(\mathcal{T}_{\mathcal{S}})\mathbf{D}(\mathbf{I}(\mathcal{T}_{\mathcal{S}}))^{\mathsf{T}}\right\vert},\quad \forall \mathcal{S} \in \bar{\mathcal{S}},\nonumber\\
&&{\mathrm{C}2:} \,\,\,& r_m\leq  {C}_m^{\mathrm{fso}}+\alpha_m{C}_m^{\mathrm{rf}},\quad\forall m \in \mathcal{M,}\nonumber\\
&&{\mathrm{C}3:} \,\,\, & \mathbf{I}(\mathcal{T})\mathbf{D}(\mathbf{I}(\mathcal{T}))^{\mathsf{T}}\succeq \mathbf{0} \quad \text{and}\quad 
\mathbf{I}(\mathcal{T})\mathbf{D}(\mathbf{I}(\mathcal{T}^{\mathsf{c}}))^{\mathsf{T}}=\mathbf{0}_{|\mathcal{T}|\times|\mathcal{T}^{\mathsf{c}}|},\quad \forall \mathcal{T} \in \bar{\mathcal{T}},
\end{IEEEeqnarray}
where $\mathcal{T}^{\mathsf{c}}$ is the complement set of $\mathcal{T}$ w.r.t. $\mathcal{N}_T$. Furthermore, $\mathbf{V}_k$, $\mathbf{W}_k$, and $\mathbf{C}(\mathcal{S})$ are constant matrices and $\bar{\mathcal{S}}$ and $\bar{\mathcal{T}}$ are constant sets defined as follows
\begin{IEEEeqnarray}{rlrl}\label{Eq:Sets}
\mathbf{V}_k&=
\begin{cases}
\sum_{j=1}^{K}P_j\mathbf{h}_{j}\mathbf{h}_{j}^{\mathsf{H}}+\sigma^2\mathbf{I}_{MN}, &\text{MMSE}\\
\sum_{j=k}^{K}P_j\mathbf{h}_{j}\mathbf{h}_{j}^{\mathsf{H}}+\sigma^2\mathbf{I}_{MN}, &\text{SIC}
\end{cases}
\quad
&\mathbf{W}_k&=
\begin{cases}
\sum_{j\neq k}^{K}P_j\mathbf{h}_{j}\mathbf{h}_{j}^{\mathsf{H}}+\sigma^2\mathbf{I}_{MN}, &\text{MMSE}\\
\sum_{j>k}^{K}P_j\mathbf{h}_{j}\mathbf{h}_{j}^{\mathsf{H}}+\sigma^2\mathbf{I}_{MN}, &\text{SIC}
\end{cases}
\nonumber\\
\mathbf{C}(\mathcal{S})&=
\begin{cases}
\mathrm{diag}\left\{\mathbf{H}(\mathcal{S})\boldsymbol{\Sigma}(\mathbf{H}(\mathcal{S}))^{\mathsf{H}}\right\}, &\text{AVQ}\\
\mathbf{H}(\mathcal{S})\boldsymbol{\Sigma}(\mathbf{H}(\mathcal{S}))^{\mathsf{H}}, &\text{RVQ and DSC}
\end{cases}
\quad
&\bar{\mathcal{S}}&=
\begin{cases}
\{\forall\mathcal{S}\subset \mathcal{M}\mid |\mathcal{S}|=1\},&\text{AVQ and RVQ}\\
\{\forall \mathcal{S}\mid\mathcal{S}\subseteq \mathcal{M}\}, &\text{DSC}
\end{cases}
\nonumber
\end{IEEEeqnarray}\vspace{-8mm}
\begin{IEEEeqnarray}{ccc}
\bar{\mathcal{T}}=
\begin{cases}
\{\forall\mathcal{T}\subset \mathcal{N}_T\mid |\mathcal{T}|=1\}, &\text{AVQ}\\
\{\forall \mathcal{T}_{\mathcal{S}}\mid\mathcal{S}\in \mathcal{M}\}, &\text{RVQ}\\
\{\forall\mathcal{T}\subseteq \mathcal{N}_T\mid |\mathcal{T}|=MN\}, &\text{DSC.}
\end{cases}
\nonumber
\end{IEEEeqnarray}
\end{prop}
\begin{IEEEproof}
The proof is provided in Appendix \ref{App:Pro}.
\end{IEEEproof}

Note that optimization problem (\ref{Eq:Unified}) is jointly non-convex in $(\boldsymbol{\alpha},\mathbf{D},\mathbf{r})$. Hence, a brute-force search may be needed for finding the globally optimal solution which entails a prohibitively high computational complexity. In addition, we note that given $(\boldsymbol{\alpha},\mathbf{r})$, (\ref{Eq:Unified}) is non-convex w.r.t. $\mathbf{D}$ since the objective function of the \textit{maximization} problem is convex instead of concave. Therefore, in the following, we  reformulate the objective function of (\ref{Eq:Unified}) such that the problem becomes bi-convex, i.e., a problem that is convex in each optimization variable assuming the other variables are fixed. This facilitates the application of ACO for finding a suboptimal solution to (\ref{Eq:Unified}).
To further reduce complexity, we transform constraint $\mathrm{C}2$ such that the optimization vector $\boldsymbol{\alpha}$ is replaced by scalar optimization variable $\alpha_0$. The optimal $\alpha_0$ can then be found with a simple one dimensional search.

\subsection{Problem Transformation} 
In this section, we present a transformation of the constraints and the objective function which allows us to efficiently tackle non-convex optimization problem (\ref{Eq:Unified}).

\subsubsection{Constraint Transformation}
Given $\mathbf{D}$, the problem is linear w.r.t. $(\boldsymbol{\alpha},\mathbf{r})$ and hence can theoretically be handled by ACO. However, the ACO algorithm fails to converge to an efficient suboptimal solution due to the following reason. For a given $\mathbf{D}=\mathbf{D}^{(i)}$ in the $i$-th iteration, the optimal  $(\boldsymbol{\alpha},\mathbf{r})=(\boldsymbol{\alpha}^{(i)},\mathbf{r}^{(i)})$ is found such that constraint $\mathrm{C}2$ holds with equality. Therefore, in the next iteration, given $(\boldsymbol{\alpha},\mathbf{r})=(\boldsymbol{\alpha}^{(i)},\mathbf{r}^{(i)})$, the optimal $\mathbf{D}=\mathbf{D}^{(i+1)}$ returns the same solution as in the previous iteration, i.e., $\mathbf{D}^{(i+1)}=\mathbf{D}^{(i)}$. In other words, the ACO algorithm will be trapped at the initial point $\mathbf{D}^{(i)}$. To overcome this challenge, we reformulate the constraints in terms of $\boldsymbol{\alpha}$ such that the reformulated problem does not exhibit the aforementioned issue. The following lemma  provides an equivalent representation of the channel capacity constraint $\mathrm{C}2$ in~(\ref{Eq:Unified}) which is useful for handling $\boldsymbol{\alpha}$.

\begin{lem}\label{LemmaA}
Constraint $\mathrm{C}2$ in (\ref{Eq:Unified}) can be written in the following equivalent form
\begin{IEEEeqnarray}{lll}\label{Eq:E_Constraint}
\widetilde{\mathrm{C}}2:\quad\sum_{\forall m\in \mathcal{S}} r_m G_m(\mathcal{S})\leq (1-\alpha_0)G(\mathcal{S})+\sum_{\forall m\in \mathcal{S}}G_m(\mathcal{S}){C}_m^{\mathrm{fso}},\,\, \forall\mathcal{S} \subseteq \mathcal{M},\quad\,\,
\end{IEEEeqnarray}
where $G_m(\mathcal{S})=\frac{\prod_{\forall m' \in \mathcal{S}}C_{m'}^{\mathrm{rf}}}{C_m^{\mathrm{rf}}}$ , $G(\mathcal{S})=\prod_{\forall m\in \mathcal{S}}{C}_m^{\mathrm{rf}}$, and $\mathcal{S}$ denotes a non-empty subset of $\mathcal{M}$.
\end{lem} 
\begin{IEEEproof}
The proof is provided in Appendix \ref{App:Lem}.
\end{IEEEproof}
%
%
\begin{remk}
The equivalence of constraint $\widetilde{\mathrm{C}}2$ in (\ref{Eq:E_Constraint}) and $\mathrm{C}2$ in (\ref{Eq:Unified}) is analogous to the equivalence of the capacity region of the MAC and the rate region achieved via time-sharing and SIC, see~\cite[Chapter~15]{Cover}. The advantage of Lemma~\ref{LemmaA} is that the $(M+1)$-dimensional optimization variable $\boldsymbol{\alpha}$ in $\mathrm{C}2$ in (\ref{Eq:Unified}) reduces to the one-dimensional optimization variable $\alpha_0$  in $\widetilde{\mathrm{C}}2$ in (\ref{Eq:E_Constraint}). This comes at the expense of increasing the number of constraints from $M$ to $2^{M}-1$.
\end{remk}
\subsubsection{Golden Section Search (GSS)}
Now, since variable $\alpha_0$ is bounded in the interval $[0,1]$, the optimal $\alpha_0^*$ can be obtained via a full search assuming the optimal $(\mathbf{D}^*,\mathbf{r}^*)$ have been already obtained for any given $\alpha_0\in[0,1]$. Note that since $\alpha_0$ is one-dimensional and bounded, a full search is feasible; however, the optimal $\alpha_0^*$ can be found even more efficiently as described in the following. In particular, assuming that the weighted sum rate is a unimodal function w.r.t. $\alpha_0$, i.e., a function that has only one optimal point in a given bounded interval~\cite{GSS}, the efficient GSS algorithm can be used to find the optimal $\alpha_0^*$. Please refer to Appendix \ref{App:Unim} for a detailed discussion of the unimodality of the weighted sum rate and the proof for the special case of $K=M=N=1$. For the general case, a rigorous proof for the unimodality of the weighted sum rate is cumbersome.  Nevertheless, our simulation studies in Section VI suggest that the weighted sum rate is in general unimodal w.r.t. $\alpha_0$. Furthermore, in the following, an intuitive justification for this property is provided. In particular, increasing $\alpha_0$ affects the weighted sum rate in two aspects, namely the RF multiple-access time interval increases and the distortion caused by the quantization also increases as the RF fronthaul time interval decreases. In other words, by increasing $\alpha_0$, the weighted sum rate first increases owing to the increasing RF multiple-access time interval, but ultimately decreases due to the decreasing fronthaul capacity and the resulting larger distortion. We note that the required number of iterations, denoted by $n_I$,  for finding $\alpha_0^*$ with an accuracy of $\epsilon$ is $n_I\approx\log \frac{1}{\epsilon}$ for the GSS and $n_I=\frac{1}{\epsilon}$ for the full search~\cite{GSS}. Hence, we employ the GSS to find the optimal $\alpha_0^*$. In the following, we  assume $\alpha_0$ is fixed and find the corresponding optimal $(\mathbf{D}^*,\mathbf{r}^*)$.

\subsubsection{Objective Function Transformation} 
The problem in (\ref{Eq:Unified}) is non-convex in $\mathbf{D}$ since the objective function of the \textit{maximization} problem is convex in $\mathbf{D}$ (instead of concave). We use the following lemma to convexify the objective function of (\ref{Eq:Unified}) w.r.t. $\mathbf{D}$.

\begin{lem}[\hspace{-0.3mm}\cite{lemma}]\label{LemmaD} For any matrix $\mathbf{X}\in \mathbb{C}^{J\times J}$ which satisfies $\mathbf{X} \succ \mathbf{0}$, the following equation holds
\begin{IEEEeqnarray}{lll}\label{Eq:Lemma}
\mathrm{log_2}|\mathbf{X}^{-1}| = \underset{\mathbf{Y} \succeq \mathbf{0}}{\mathrm{max}}\,\,\mathrm{log_2}|\mathbf{Y}|-\dfrac{1}{\ln(2)}\mathrm{Tr}(\mathbf{Y}\mathbf{X})+\dfrac{J}{\ln(2)},
\end{IEEEeqnarray}
where the optimal solution of the right-hand side of (\ref{Eq:Lemma}) is given by $\mathbf{Y^*}=\mathbf{X}^{-1}$.
\end{lem}

Defining $\mathbf{X}_k=\mathbf{W}_k+\mathbf{D}$ and $\mathbf{Y}_k=\mathbf{B}_k$, where $\mathbf{B}_k$ is a new auxiliary optimization matrix, and applying Lemma~\ref{LemmaD} to the objective function of (\ref{Eq:Unified}) and replacing the original constraint ${\mathrm{C}}2$ with the equivalent constraint $\widetilde{\mathrm{C}}2$ from Lemma~\ref{LemmaA}, we reformulate optimization problem (\ref{Eq:Unified}) as follows
\begin{IEEEeqnarray}{llll}\label{Eq:TOTAL}
 &\underset{\mathbf{r},\mathbf{B}_k\succeq \mathbf{0},\forall k}{\underset{\alpha_0\in[0,1] ,\mathbf{D},}{\mathrm{maximize}}}\,\,&T=&\alpha_0 W^{\mathrm{rf}}\sum_{k=1}^{K} \mu_{k}\mathrm{log_2}\left\vert\mathbf{V}_k+\mathbf{D}\right\vert
+\mu_{k}\mathrm{log_2}\left\vert \mathbf{B}_k\right\vert -\dfrac{\mu_{k}}{\ln(2)}\mathrm{Tr}\left(\mathbf{B}_k\left(\mathbf{W}_k+\mathbf{D}\right)\right)\\
&\mathrm{subject\,\, to} \,\,\,  &{\mathrm{C}1}:\,\,&\sum_{m\in\mathcal{S}}r_m \geq \alpha_0 f_s \mathrm{log_2}\dfrac{\left\vert \mathbf{C}(\mathcal{S})+\mathbf{I}(\mathcal{T}_{\mathcal{S}})\mathbf{D}(\mathbf{I}(\mathcal{T}_{\mathcal{S}}))^{\mathsf{T}}+\sigma^2\mathbf{I}_{|\mathcal{S}|N}\right\vert}{\left\vert \mathbf{I}(\mathcal{T}_{\mathcal{S}})\mathbf{D}(\mathbf{I}(\mathcal{T}_{\mathcal{S}}))^{\mathsf{T}}\right\vert},\quad \forall \mathcal{S} \in \bar{\mathcal{S}},\nonumber\\
&&{\widetilde{\mathrm{C}}2:} \,\,\,& \sum_{\forall m\in \mathcal{S}} r_m G_m(\mathcal{S})\leq (1-\alpha_0)G(\mathcal{S})+\sum_{\forall m\in \mathcal{S}}G_m(\mathcal{S}){C}_m^{\mathrm{fso}},\,\, \forall\mathcal{S} \subseteq \mathcal{M},\nonumber\\
&&{\mathrm{C}3:} \,\,\, & \mathbf{I}(\mathcal{T})\mathbf{D}(\mathbf{I}(\mathcal{T}))^{\mathsf{T}}\succeq \mathbf{0} \quad \text{and}\quad 
\mathbf{I}(\mathcal{T})\mathbf{D}(\mathbf{I}(\mathcal{T}^{\mathsf{c}}))^{\mathsf{T}}=\mathbf{0}_{|\mathcal{T}|\times|\mathcal{T}^{\mathsf{c}}|},\,\, \forall \mathcal{T} \in \bar{\mathcal{T}}.\quad\nonumber
\end{IEEEeqnarray}

Although, for a given $\alpha_0$, optimization problem (\ref{Eq:TOTAL}) is still jointly non-convex in $(\mathbf{D},\mathbf{r},\mathbf{B}_k)$, the problem is convex w.r.t. the individual variables. This allows the use of ACO to find a suboptimal solution of the problem in terms of $(\mathbf{D},\mathbf{r},\mathbf{B}_k)$ for any given $\alpha_0$. In particular, given $(\mathbf{r},\mathbf{B}_k)$, the problem is convex w.r.t. $\mathbf{D}$ and can be solved using standard semi-definite programming (SDP) solvers~\cite{cvx}. Moreover, given $(\mathbf{D},\mathbf{r})$, the problem is convex w.r.t. $\mathbf{B}_k$ and has the following optimal closed-form solution based on Lemma~\ref{LemmaD}\vspace{-3mm}
\begin{IEEEeqnarray}{lll}\label{Eq:B}
\mathbf{B}_k^*=\left(\mathbf{W}_k+\mathbf{D}\right)^{-1}.
\end{IEEEeqnarray}
 
 In the following subsection, we propose a nested loop algorithm which exploits Lemma~\ref{LemmaA} and Lemma~\ref{LemmaD}.

\subsection{Proposed Algorithm}
We propose an algorithm consisting of an outer loop, i.e., Algorithm~1, to find $\alpha_0^*$ based on GSS, and an inner loop, i.e., Algorithm~2, to find $(\mathbf{D}^*,\mathbf{r}^*,\mathbf{B}_k^*)$ for a given $\alpha_0$ based on ACO, respectively.  In the following, we describe the outer and inner loops in detail.
\begin{figure}
\begin{algorithm}[H] \label{Alg:Outer Loop}
\caption{GSS-Outer Loop}
\begin{algorithmic}[1] 
\STATE \textbf{input} $\epsilon$, $\alpha_0^{\mathrm{min}}=0$, and $\alpha_0^{\mathrm{max}}=1$.
\REPEAT 
\STATE Update $\alpha_0^{(1)}$ as in (\ref{Eq:GSS_Points}), obtain $\mathbf{D}^{(1)}$ from \textbf{{Algorithm~2}}, and compute $T^{(1)}$ from  (\ref{Eq:TOTAL}).
\STATE Update $\alpha_0^{(2)}$ as in (\ref{Eq:GSS_Points}), obtain $\mathbf{D}^{(2)}$ from \textbf{{Algorithm~2}}, and compute $T^{(2)}$ from  (\ref{Eq:TOTAL}).
\IF{$T^{(1)}\geq T^{(2)}$} 
\STATE Update $\alpha_0^{\mathrm{max}}=\alpha_0^{(2)}$,
\ELSIF{$T^{(1)}< T^{(2)}$}
 \STATE Update $\alpha_0^{\mathrm{min}}=\alpha_0^{(1)}$.
\ENDIF 
\UNTIL{$|\alpha_{0}^{\mathrm{max}}- \alpha_{0}^{\mathrm{min}}| \leq \epsilon$}
\STATE $\alpha_0^{*}\gets (\alpha_{0}^{\mathrm{min}}+ \alpha_{0}^{\mathrm{max}})/2$  and obtain $\mathbf{D}^{*}$ and $\mathbf{r}^{*}$ from \textbf{Algorithm~2}.
\STATE \textbf{output} $\alpha_0^{*}$, $\mathbf{D}^{*}$, and $\mathbf{r}^{*}$.
\end{algorithmic}
\end{algorithm}
\vspace{-1cm}
\begin{algorithm}[H]  \label{Alg:Inner Loop}
\caption{ACO (\textcolor{blue}{\textit{M-ACO}})-Inner Loop}
\begin{algorithmic}[1] 
\STATE \textbf{input} $\varepsilon$, $n^{\max}$, $i=0$, $T^{\{0\}}=0$, ${\mathbf{D}}^{\{0\}}=d_0 \mathbf{I}_{MN}$, and $\alpha_{0}$.
\REPEAT
\STATE $i \gets i+1$.
\STATE Given ${\mathbf{D}}^{\{i-1\}}$, update $\mathbf{B}^{\{i-1\}}_k,\forall k$ (\textcolor{blue}{$\mathbf{\textit{B}}^{\textit{\{i\}}}_k,\forall k$ \textit{and} $\mathbf{\textit{A}}^{\textit{\{i\}}}$}) from the closed-form expression (\textcolor{blue}{\textit{expressions}})  given in (\ref{Eq:B}) (\textcolor{blue}{\textit{(\ref{Eq:B}) and (\ref{Eq:A_m}), respectively}}).
\STATE Given $\mathbf{B}^{\{i\}}_k$ and $\alpha_0$, update ${\mathbf{D}}^{\{i\}}$ and ${\mathbf{r}}^{\{i\}}$ by solving SDP problem (\ref{Eq:TOTAL}) (\textcolor{blue}{\textit{(\ref{Eq:TOTAL_CVX})}}).
\STATE Given $\alpha_0$, ${\mathbf{D}}^{\{i\}}$, and $\mathbf{B}^{\{i\}}_k$, compute $T^{\{i\}}$ from the objective function in (\ref{Eq:TOTAL}) (\textcolor{blue}{\textit{(\ref{Eq:TOTAL_CVX})}}).
\UNTIL{$|T^{\{i\}}-T^{\{i-1\}}|\leq \varepsilon$ or $i\geq n^{\max}$}
\STATE \textbf{output} $\mathbf{D}={\mathbf{D}}^{\{i\}}$  and $\mathbf{r}=\mathbf{r}^{\{i\}}$.
\end{algorithmic}
\end{algorithm}
\vspace{-1cm}
\end{figure}

\textit{Outer Loop (Algortihm~1):}
In this loop, we employ the iterative GSS algorithm to maximize the weighted sum rate w.r.t. $\alpha_0$. Suppose $[\alpha_0^{\mathrm{min}},\alpha_0^{\mathrm{max}}]$ is the search interval in a given iteration. In each iteration, the GSS algorithm requires the evaluation of the weighted sum rate at the intermediate points $(\alpha_0^{(1)},\alpha_0^{(2)})$ using the inner loop, i.e., Algorithm~2. Then, it compares the values of the objective function at $\alpha_0^{(1)}$ and $\alpha_0^{(2)}$ denoted by $T^{(1)}$ and $ T^{(2)}$, respectively, and updates the search interval for the next iteration as follows~\cite{GSS}
\begin{IEEEeqnarray}{llll}\label{Eq:GSS_update}
\text{Search Interval}=
\begin{cases}
[\alpha_0^{\min},\alpha_0^{(2)}]\qquad \mathrm{if}\,\, T^{(1)}\geq T^{(2)},\\
[\alpha_0^{(1)},\alpha_0^{\max}]\qquad \mathrm{if}\,\, T^{(1)}< T^{(2)},
\end{cases}
\end{IEEEeqnarray}
where  $(\alpha_0^{(1)},\alpha_0^{(2)})$ are obtained as 
\begin{IEEEeqnarray}{llll}\label{Eq:GSS_Points}
(\alpha_0^{(1)},\alpha_0^{(2)})=\left(\alpha_0^{\mathrm{min}}+\rho\Delta\alpha,\alpha_0^{\max}-\rho\Delta\alpha\right).
\end{IEEEeqnarray}
Here, $\Delta\alpha=\alpha_0^{\mathrm{max}}-\alpha_0^{\mathrm{min}}$ and  $\rho=1-\frac{1}{\phi}$ with the so-called golden ratio $\phi=(1+\sqrt{5})/{2}\approx 1.61803$~\cite{GSS}. The GSS algorithm is summarized in Algorithm~1 for the problem considered in this paper. In Algorithm~1, $\epsilon$ is a small positive number which is used in line~10 to terminate the iteration if the desired accuracy of the GSS algorithm is achieved.

\textit{Inner Loop (Algortihm~2):} 
For the inner loop, $\alpha_0$ is provided by the outer loop. Here, we employ ACO to find a stationary point of the problem w.r.t. $\mathbf{D}$, $\mathbf{r}$, and $\mathbf{B}_k$ for the fixed $\alpha_0$. The proposed ACO method is concisely given in Algorithm~2 where $\varepsilon$ is a small positive number and $n^{\max}$ is the maximum number of iterations that are used in the termination condition in line~7. In the next section, we provide a modified ACO (M-ACO) method, which is easier to implement in popular numerical solvers such as CVX and also included in Algorithm~2 in blue italic font. 

Once $\mathbf{r}^*$ is known, the optimal RF time allocation variable for fronthaul link RU$_m$-CU, $\alpha_m^*$, is obtained~as\vspace{-3mm}
\begin{IEEEeqnarray}{lll}\label{Eq:alpha_m}
\alpha_m^*=\frac{r_m^*-C_m^{\mathrm{fso}}}{C_m^{\mathrm{rf}}}, \,\,\forall m\in \mathcal{M}.
\end{IEEEeqnarray}


\subsection{Implementation in CVX}
 Although for given $\alpha_0$, $\mathbf{r}$, and $\mathbf{B}_k,\forall k$, optimization problem (\ref{Eq:TOTAL}) is convex in $\mathbf{D}$, its solution  using popular numerical solvers, such as CVX \cite{cvx}, can be challenging. More specifically, the implementation of line~5 of Algorithm~2 in CVX is not directly possible since the current version of CVX, i.e., CVX 2.1, does not have a built-in convex function that can directly handle the right-hand side of $\mathrm{C}1$ in (\ref{Eq:TOTAL}) \cite{cvx}. In the following, we propose a transformation of ${\mathrm{C}1}$ denoted by $\widetilde{{\mathrm{C}}}1$ to address this issue. Defining $\mathbf{X}(\mathcal{S})=\mathbf{C}(\mathcal{S})+\mathbf{I}(\mathcal{T}_{\mathcal{S}})\mathbf{D}(\mathbf{I}(\mathcal{T}_{\mathcal{S}}))^{\mathsf{T}}+\sigma^2\mathbf{I}_{|\mathcal{S}|N}$ and $\mathbf{Y}(\mathcal{S})=\mathbf{A}(\mathcal{S})$, where $\mathbf{A}(\mathcal{S})$ is a new auxiliary optimization matrix, and using (\ref{Eq:Lemma}) in Lemma~\ref{LemmaD}, the following inequality holds for the right-hand side of $\mathrm{C}1$ in (\ref{Eq:TOTAL})
\begin{IEEEeqnarray}{lll}\label{Eq:R_ub}
\mathrm{log_2}\dfrac{\left\vert \mathbf{C}(\mathcal{S})+\mathbf{I}(\mathcal{T}_{\mathcal{S}})\mathbf{D}(\mathbf{I}(\mathcal{T}_{\mathcal{S}}))^{\mathsf{T}}+\sigma^2\mathbf{I}_{|\mathcal{S}|N}\right\vert}{\left\vert \mathbf{I}(\mathcal{T}_{\mathcal{S}})\mathbf{D}(\mathbf{I}(\mathcal{T}_{\mathcal{S}}))^{\mathsf{T}}\right\vert}\nonumber\\
\qquad\leq -\mathrm{log_2}|\mathbf{A}(\mathcal{S})|
+\dfrac{1}{\ln(2)}\mathrm{Tr}\left(\mathbf{A}(\mathcal{S})\left(\mathbf{C}(\mathcal{S})+\mathbf{I}(\mathcal{T}_{\mathcal{S}})\mathbf{D}(\mathbf{I}(\mathcal{T}_{\mathcal{S}}))^{\mathsf{T}}+\sigma^2\mathbf{I}_{|\mathcal{S}|N}\right)\right)\nonumber\\
\qquad\quad-\dfrac{|\mathcal{S}|N}{\ln(2)}-\mathrm{log_2}\left\vert \mathbf{I}(\mathcal{T}_{\mathcal{S}})\mathbf{D}(\mathbf{I}(\mathcal{T}_{\mathcal{S}}))^{\mathsf{T}}\right\vert\triangleq R^{\mathrm{ub}},\quad\forall \mathbf{A}(\mathcal{S})\succeq \mathbf{0}.
\end{IEEEeqnarray}
 In (\ref{Eq:R_ub}), equality holds if 
\begin{IEEEeqnarray}{lll}\label{Eq:A_m}
\mathbf{A}^*(\mathcal{S})=\left(\mathbf{C}(\mathcal{S})+\mathbf{I}(\mathcal{T}_{\mathcal{S}})\mathbf{D}(\mathbf{I}(\mathcal{T}_{\mathcal{S}}))^{\mathsf{T}}+\sigma^2\mathbf{I}_{|\mathcal{S}|N}\right)^{-1},\quad \forall \mathcal{S} \in \bar{\mathcal{S}}.
\end{IEEEeqnarray}
Substituting (\ref{Eq:R_ub}) into ${\mathrm{C}1}$ in (\ref{Eq:TOTAL}), we obtain $\widetilde{{\mathrm{C}}}1$ as follows 
\begin{IEEEeqnarray}{lll}\label{Eq:C1_tilde}
\widetilde{{\mathrm{C}}}1:\sum_{m\in\mathcal{S}}r_m \geq \alpha_0 f_s R^{\mathrm{ub}},
\end{IEEEeqnarray}
which is convex in $\mathbf{D}$ if $\mathbf{A}(\mathcal{S})$ is fixed and vice versa. We note that the feasible set of ${\mathrm{C}1}$ w.r.t. $(\alpha_0,\mathbf{D},\mathbf{r})$ is identical to that of $\widetilde{{\mathrm{C}}}1$ w.r.t. $(\alpha_0,\mathbf{D},\mathbf{r},\mathbf{A}(\mathcal{S}))$. In fact, since for any $\mathbf{A}(\mathcal{S})\succeq \mathbf{0}$ (\ref{Eq:R_ub}) holds, the feasible set of $\widetilde{{\mathrm{C}}}1$ cannot be larger than that of ${\mathrm{C}1}$. On the other hand, since $\mathbf{A}(\mathcal{S})$ contains $\mathbf{A}(\mathcal{S})=\mathbf{A}^*(\mathcal{S})$ as a special case, the feasible set of $\widetilde{{\mathrm{C}}}1$ cannot be smaller than that of ${\mathrm{C}1}$ either. Hence, the feasible sets of ${\mathrm{C}1}$ and $\widetilde{{\mathrm{C}}}1$ are identical. Therefore, we can rewrite problem (\ref{Eq:TOTAL}) in the following equivalent form
\begin{IEEEeqnarray}{llll}\label{Eq:TOTAL_CVX}
 &\underset{\mathbf{B}_k\succeq \mathbf{0},\forall k,\mathbf{A}(\mathcal{S})\succeq \mathbf{0},\forall \mathcal{S}}{\underset{\alpha_0\in[0,1],\mathbf{D},\mathbf{r}}{\mathrm{maximize}}}\,\,&T\\
&\mathrm{subject\,\, to} \,\,\,  &\widetilde{{\mathrm{C}}}1\text{,}\,\,\,\widetilde{{\mathrm{C}}}2,\,\,\,\text{and}\,\,\, {\mathrm{C}3},\nonumber 
\end{IEEEeqnarray}
where $T$, $\widetilde{{\mathrm{C}}}2$, and ${\mathrm{C}3}$ are given in (\ref{Eq:TOTAL}) and $\widetilde{{\mathrm{C}}}1$ is given in (\ref{Eq:C1_tilde}).

Based on (\ref{Eq:TOTAL_CVX}), an M-ACO algorithm w.r.t. $(\mathbf{D},\mathbf{r},\mathbf{B}_k,\mathbf{A}(\mathcal{S}))$ can be developed. The corresponding changes are provided in blue italic font in Algorithm~2.
\begin{remk}
In summary, we note that the global optimal solutions of the three non-convex optimization problems (\ref{Eq:Unified}), (\ref{Eq:TOTAL}), and (\ref{Eq:TOTAL_CVX}) lead to identical values for the weighted sum rate. However, unlike original problem (\ref{Eq:Unified}),  problem (\ref{Eq:TOTAL}) can be solved suboptimally using ACO, and problem (\ref{Eq:TOTAL_CVX}) can be solved suboptimally using ACO via CVX. The obtained suboptimal solutions are local optima of original problem (\ref{Eq:Unified}) because of the convergence properties of ACO, see e.g.~\cite{Kwan_SDP_Complexity,ACO_Local_Optimum}.
Therefore, we use (\ref{Eq:TOTAL_CVX}) to generate the simulation results shown in Section VI.
\end{remk}

\section{Complexity Analysis}
In this section, we first analyze the computational complexity of Algorithms~1 and 2. Subsequently, we characterize the complexity of the quantization  and detection schemes adopted in this paper for online transmission. 

\subsection{Algorithms 1 and 2}
In the complexity analysis presented below, we only consider the operations that entail the highest computational complexity. In particular, calculating the matrix inversion in Step 4 and solving the SDP problem in Step 5 of Algorithm~2 dominate the overall complexity of both Algorithms~1 and 2 and are discussed in the following.

\textit{Step~4:} The computational complexity of inverting a matrix of size $n\times n$  is $\mathcal{O}(n^3)$ \cite{Matrix,Hunger_FlOPs}. The ACO in Algorithm~2 includes a matrix inversion for finding $\mathbf{B}_k$ for $K$ MUs which entails a complexity of $\mathcal{O}(KM^3N^3)$. For M-ACO, matrix inversions are needed to find both $\mathbf{B}_k$ and $\mathbf{A}(\mathcal{S})$. The complexity of computing $\mathbf{A}(\mathcal{S})$ depends on the type of quantization employed at the RUs, cf. (\ref{Eq:A_m}), and is $\mathcal{O}(MN)$ for AVQ since $\mathbf{A}(\mathcal{S})$ is diagonal, $\mathcal{O}(MN^3)$ for RVQ since $\mathbf{A}(\mathcal{S})$ is block diagonal, and $\mathcal{O}\left( \sum_{i=1}^{M}\binom{M}{i} \cdot (iN)^3\right)\approx \mathcal{O}(2^{M}M^3N^3)$ for DSC since $\mathbf{A}(\mathcal{S})$ is a general matrix and has to be computed for $\forall\mathcal{S}	\subseteq\mathcal{M}$. Therefore, the overall complexity of computing $\mathbf{A}(\mathcal{S})$ and $\mathbf{B}_k$  in Step 4 of M-ACO is  $\mathcal{O}\left(KM^3N^3\right)$ for AVQ and RVQ and $\mathcal{O}\left(\left(K+2^M\right)M^3N^3\right)$ for DSC.

\textit{Step 5:} The computational complexity required per iteration for solving an SDP with a numerical convex program solver is given by $\mathcal{O}(mn^3+m^2n^2+m^3)$ \cite{SDPComplexity,Kwan_SDP_Complexity}, where $m$ and $n$ denote the number of semidefinite  cone constraints and the dimension of the semidefinite cone, respectively.
Moreover, the number of iterations needed to solve the SDP problem with an accuracy of $\xi_{\mathrm{sdp}}$ is on the order of $\mathcal{O}(\sqrt{n}\log\frac{1}{\xi_{\mathrm{sdp}}})$ iterations \cite{SDPComplexity,Kwan_SDP_Complexity}. This leads to a complexity order of $\mathcal{O}(mn^{3.5}+m^2n^{2.5}+m^3n^{0.5})$. 
For AVQ, we do not have semidefinite cones; hence the complexity of Step~4 is higher than the complexity of Step~5. Moreover, for RVQ, we have $m=M$ and $n=N$. Hence, the complexity of solving one SDP is $\mathcal{O}(MN^{3.5}+M^2N^{2.5}+M^3N^{0.5})$. For DSC, $m=1$ and $n=MN$, and the corresponding complexity for solving one SDP is $\mathcal{O}(M^{3.5}N^{3.5})$. 

\textit{Overall:}  The overall complexity of the proposed algorithms is limited by the complexity of Steps 4 and 5 of Algorithm~2. Moreover, Steps 4 and 5 are repeated at most $n_{\max}$ times for Algorithm~2 to converge and $\log\frac{1}{\epsilon}$ times for Algorithm 1 to converge. Therefore, the overall complexity order of the proposed algorithm is given by
\begin{IEEEeqnarray}{lll}\label{Eq:Complexity}
\begin{cases}
\mathcal{O}\left(KM^3N^3n^{\max}\log\dfrac{1}{\epsilon}\right)\quad &\text{ACO/M-ACO algorithm with AVQ},\\
\mathcal{O}\left((KM^3N^3+MN^{3.5})n^{\max}\log\dfrac{1}{\epsilon}\right)\quad &\text{ACO/M-ACO algorithm with RVQ},\\
\mathcal{O}\left((KM^3N^3+M^{3.5}N^{3.5})n^{\max}\log\dfrac{1}{\epsilon}\right)\quad &\text{ACO algorithm with DSC},\\
\mathcal{O}\left(((K+2^M)M^3N^3+M^{3.5}N^{3.5})n^{\max}\log\dfrac{1}{\epsilon}\right)\quad &\text{M-ACO algorithm with DSC.}
\end{cases}
\end{IEEEeqnarray}
As can be seen from (\ref{Eq:Complexity}), the complexities of both ACO and M-ACO grow linearly w.r.t. the number of MUs, $K$, for all considered quantization and detection schemes. However, the growth speed of the complexity of ACO/M-ACO in terms of the number of RUs, $M$, and their number of antennas, $N$, depends on the adopted quantization scheme. For instance, the complexity of ACO/M-ACO  in terms of the number of antennas, $N$, for AVQ is cubic while it is on the order of $N^{3.5}$ for RVQ and DSC.

\subsection{Online Transmission}
After determining the optimal strategy with Algorithms~1 and 2, online transmissions based on the adopted quantization and detection schemes starts. In the following, we characterize the complexity of the adopted quantization and detection schemes for online transmission. 

\textit{Quantization Schemes:} The complexity of the adopted quantization schemes is affected by both the encoding at the RUs and the decoding at the CU. The estimated complexity depends on the size of the quantization codebooks and is summarized as follows \cite{Cover} 
\begin{IEEEeqnarray}{lll}\label{Eq:Complexity_Q}
\begin{cases}
\sum_{m=1}^M\sum_{n=1}^N 2^{I(y_{m,n};\hat{y}_{m,n})}\sim\mathcal{O}\left(MN\right) &\text{AVQ (Encoding/Decoding)},\\
\sum_{m=1}^M N2^{I(\mathbf{y}_{m};\hat{\mathbf{y}}_{m})}\sim\mathcal{O}\left(MN2^N\right) &\text{RVQ (Encoding/Decoding) and DSC (Encoding)},\\
 MN2^{I(\mathbf{y};\hat{\mathbf{y}})}\sim\mathcal{O}\left(MN2^{MN}\right) &\text{DSC (Decoding)}.
\end{cases}\quad
\end{IEEEeqnarray}
The encoding complexity of DSC is on the same order as that of RVQ, although it may involve further signaling overhead as the CU has to inform the RUs which quantization codebooks they should adopt. Nevertheless, the bottleneck of the overall complexity of DSC is on the decoder side.
Note that the complexity of the quantization schemes is not a function of the number of MUs $K$. Moreover, as expected, DSC is more complex than RVQ and AVQ, and RVQ is more complex than AVQ.

\textit{Detection Schemes:} The complexity of the detection schemes is dominated by the matrix inversion operation required for calculating the linear filter in (\ref{Eq:MMSE Filter}) \cite{Shahram_Compexity}. Note that the SINR-maximizing linear filter in (\ref{Eq:MMSE Filter}) involves the inversion of an $MN\times MN$ matrix that has to be performed for each MU separately. In the following, we show that (\ref{Eq:MMSE Filter}) can be computed much more efficiently. In particular, the linear filter based on the mean square error (MSE) metric is obtained as
\begin{IEEEeqnarray}{lll}\label{Eq:m_MMSE}
\mathbf{m}^{\mathrm{mmse}}_k=\underset{\mathbf{m}_k}{\mathrm{argmin}}\,\, \mathbbmss{E}\{|\mathbf{m}^{\mathsf{H}}_k\hat{\mathbf{y}}-x_k|^2\}=P_k\left(\mathbf{H}\boldsymbol{\Sigma}\mathbf{H}^{\mathsf{H}}+\mathbf{D}+\sigma^2\mathbf{I}_{MN}\right)^{-1}\mathbf{h}_k,
\end{IEEEeqnarray}
which requires the inversion of the same  matrix for all MUs. We note that although $\mathbf{m}^{\mathrm{mmse}}_k$ and $\mathbf{m}^{\mathrm{sinr}}_k$ are not identical, they yield the same SINR, $\gamma_k^{\mathrm{opt}}$, given in (\ref{Eq:MMSE SINR}). The computational complexity of the matrix inversion in (\ref{Eq:m_MMSE}) is $\mathcal{O}(M^3N^3)$ \cite{Hunger_FlOPs,Shahram_Compexity}. In the following, we show that for the practical case of $K\leq MN$, the computational complexity can be further reduced for AVQ and RVQ. In particular, using the Sherman-Morrison-Woodbury formula \cite{Horn_Matrix},
(\ref{Eq:m_MMSE}) is transformed to 
\begin{IEEEeqnarray}{lll}\label{Eq:Transform}
\mathbf{m}^{\mathrm{mmse}}_k=&P_k\Big(\bar{\mathbf{D}}^{-1}-\bar{\mathbf{D}}^{-1}\mathbf{H} \left(\boldsymbol{\Sigma}^{-1}+\mathbf{H}^{\mathsf{H}}\bar{\mathbf{D}}^{-1}\mathbf{H}\right)^{-1}\mathbf{H}^{\mathsf{H}}\bar{\mathbf{D}}^{-1}\Big)\mathbf{h}_k,
\end{IEEEeqnarray}
where $\bar{\mathbf{D}}=\mathbf{D}+\sigma^2\mathbf{I}_{MN}$. Considering that $\mathbf{D}$ is diagonal and block diagonal for AVQ and RVQ, respectively, the complexity of calculating the MMSE filter in (\ref{Eq:Transform}) is $\mathcal{O}\left(K^3+MN\right)$ for AVQ and $\mathcal{O}\left(K^3+MN^3\right)$ for RVQ. Therefore, the overall complexity of the linear MMSE detector for the considered quantization schemes is obtained~as
\begin{IEEEeqnarray}{lll}\label{Eq:Complexity_MMSE}
\begin{cases}
\mathcal{O}\left(K^3+MN\right) &\text{AVQ },\\
\mathcal{O}\left(K^3+MN^3\right) &\text{RVQ},\\
\mathcal{O}\left(M^3N^3\right) &\text{DSC}.
\end{cases}\quad
\end{IEEEeqnarray}


Recall that the SIC receiver is implemented by applying MMSE decision-feedback equalization to the signals received at the CU, cf. Remark~1.
Therefore,  the SIC receiver also requires an extra Cholesky factorization of the following $K\times K$ matrix \cite{MIMO}
\begin{IEEEeqnarray}{lll}\label{Eq:Cholesky}
\mathbbmss{E}\{\mathbf{x}\mathbf{x}^{\mathsf{H}}\}
-\mathbbmss{E}\{\mathbf{x}\hat{\mathbf{y}}^{\mathsf{H}}\}
\left(\mathbbmss{E}\{\hat{\mathbf{y}}\hat{\mathbf{y}}^{\mathsf{H}}\}\right)^{-1}
\mathbbmss{E}\{\hat{\mathbf{y}}\mathbf{x}^{\mathsf{H}}\}
= \boldsymbol{\Sigma} - \boldsymbol{\Sigma}\mathbf{H}^{\mathsf{H}}
\left(\mathbf{H}\boldsymbol{\Sigma}\mathbf{H}^{\mathsf{H}}+\mathbf{D}+\sigma^2\mathbf{I}_{MN}\right)^{-1} \mathbf{H}\boldsymbol{\Sigma}^{\mathsf{H}}. \quad
\end{IEEEeqnarray}
In general, Cholesky factorization of an $n\times n$ matrix entails a complexity of $\mathcal{O}(n^3)$ \cite{Matrix,Hunger_FlOPs}. Therefore, in comparison with the linear MMSE detector, SIC entails an additional complexity of $\mathcal{O}(K^3)$. 

From (\ref{Eq:Complexity_MMSE}), we observe that the complexities of the considered detection schemes depend on the adopted quantization scheme. In particular, the less complex the quantization scheme is, the less complex the corresponding MMSE or SIC detector is. Moreover,  the complexity order of the SIC detector is higher than that of the MMSE detector only in terms of the number of MUs, $K$.

\section{Simulation Results}
In the following, we first present the simulation setup and subsequently provide simulation results to evaluate the performance and complexity of the adopted quantization and detection schemes.

\subsection{Simulation Setup}
 We assume Rayleigh, Rician, and Gamma-Gamma (GGamma) fading for the RF multiple-access, RF fronthaul, and FSO channels, respectively \cite{MyTCOM,Schober}.
  Let $\beta\in\{h,f\}$ denote an RF channel gain which is given by $\beta=\sqrt{\bar{\beta}}\tilde{\beta}$, where $\bar{\beta}$ and $\tilde{\beta}$ are the average gain and the small-scale fading coefficient of the RF link, respectively. Moreover, $h$ and $f$ represent the RF access and the RF fronthaul links, respectively. The RF parameters for the access and fronthaul links are modelled as
\begin{IEEEeqnarray}{rll}\label{Eq:RF_LinkModel}
\begin{cases}
\bar{\beta}=\Bigg[\dfrac{\lambda^{\mathrm{rf}}\sqrt{G_{\mathrm{t}}^{\mathrm{Tx}}G_{\mathrm{r}}^{\mathrm{Rx}}}}{4 \pi d^{\mathrm{rf}}_{\mathrm{ref}}}\Bigg]^2\times \Bigg[\dfrac{d^{\mathrm{rf}}_{\mathrm{ref}}}{d}\Bigg]^{\nu}\\
|\tilde{\beta}|  \sim \mathrm{Rice}(\Omega,\Psi).
\end{cases}
\end{IEEEeqnarray}
Here, $\lambda^{\mathrm{rf}}$ is the wavelength of the RF signal, $G_{\mathrm{t}}^{\mathrm{Tx}}$ and $G_{\mathrm{r}}^{\mathrm{Rx}}$ are the transmit and receive RF antenna gains, respectively, where $\mathrm{t}\in\{\mathrm{MU},\mathrm{RU}\}$ and $\mathrm{r}\in\{\mathrm{RU},\mathrm{CU}\}$ denote the RF transmitter and receiver for respectively access and fronthaul.  $d^{\mathrm{rf}}_{\mathrm{ref}}$ is a reference distance for the antenna far-field, and  $d\in\{d^{\mathrm{ac}},d^{\mathrm{fr}}\}$ is the distance between RF transmitter and receiver, where $d^{\mathrm{ac}}$ and $d^{\mathrm{fr}}$ denote the access and fronthaul distances, respectively, and  $\nu$ is the path-loss exponent of the RF links. Parameters $\Omega$ and $\Psi$ of the Rice distribution denote the ratios between the power in the direct path and the power in the scattered paths to the total power in both paths, respectively. Note that for Rayleigh fading, $\Omega=0$ holds. Similarly, the distance-dependent FSO channel model is given by 
\begin{IEEEeqnarray}{rll}\label{Eq:FSO_LinkModel}
\begin{cases}
\bar{g}=\mathit{R}\left[ \mathrm{erf}\left( \dfrac{\sqrt{\pi}r}{\sqrt{2}\phi d^{\mathrm{fr}}}\right) \right] ^2\times 10^{-\kappa d^{\mathrm{fr}}/10}\\
\tilde{g}\sim \mathrm{GGamma}(\Theta,\Phi),
\end{cases}
\end{IEEEeqnarray}
where $g=\bar{g}\tilde{g}$ is the FSO channel gain, and $\bar{g}$ and $\tilde{g}$ are the average gain and the small-scale fading gain of the FSO link, respectively. Moreover, $\mathit{R}$ denotes the responsivity of the PD, $r$ is the aperture radius, $\phi$ is the divergence angle of the beam, and $\kappa$  is the weather-dependent attenuation factor of the FSO links. Parameters $\Theta$ and $\Phi$ of the GGamma distribution depend on physical parameters such as the wavelength $\lambda^{\mathrm{fso}}$ and the weather-dependent index of refraction structure parameter $C_n^2$, cf. \cite[Eqs. (3) and (4)]{Schober}. 
\begin{table}
\label{Table:Parameter}
\caption{Simulation Parameters~\cite{MyTCOM,FSO_Vahid}.\vspace{-0.4cm}} 
  \begin{minipage}[b]{0.6\textwidth}
\begin{center}
\scalebox{0.65}
{
\begin{tabular}{|| c | c  | c ||}
  \hline
   \multicolumn{3}{||c||}{\textbf{RF Link}}\\ \hline \hline    
 Symbol & Definition & Value \\ \hline \hline
 $d^{\mathrm{ac}}$ & Distance between the MUs and the RUs & $100$ m \\ \hline
  $d^{\mathrm{fr}}$ & Distance between the RUs and the CU & $500$ m \\ \hline
   $d^{\mathrm{rf}}_{\mathrm{ref}}$ & Reference distance of the RF link & $5$ m \\ \hline
   $P_{k}$ & Transmit power of MU$_k$ & $16$ dBm \\ \hline 
   $\bar{P}_{m}$ & Transmit power of RU$_m$ & $33$ dBm \\ \hline  
 $(G^{\mathrm{Tx}}_{\mathrm{MU}}, G^{\mathrm{Rx}}_{\mathrm{RU}})$  & Antenna gains for the RF  multiple-access link & $(0,10)$ dBi \\ \hline
   $(G^{\mathrm{Tx}}_{\mathrm{RU}}, G^{\mathrm{Rx}}_{\mathrm{CU}})$   & Antenna gains for  RF fronthaul link & $(10,10)$ dBi \\ \hline
  $N_{0}$ & Noise power spectral density & $-114$ dBm/MHz \\ \hline 
  $N_{F}$ & Noise figure at the RF receivers & $5$ dB \\ \hline
   $\lambda^{\mathrm{rf}}$ & Wavelength of RF signal & $85.7$ mm  \\ \hline 
      $W^{\mathrm{rf}}$ & Bandwidth of RF signal & $40$ MHz \\ \hline 
            $f_s$ & Sampling frequency & $40$ MHz \\ \hline 
     $\Omega$  & Rician fading factor & $6$ dB  \\ \hline  
    $\nu$ &  RF path-loss exponent & $3.5$ \\ \hline\hline 
\end{tabular}
}
\end{center}
  \end{minipage}
    \begin{minipage}[b]{0.3\textwidth}    
\hspace{-1cm}\vspace{-0.83cm}
\scalebox{0.65}
{
\begin{tabular}{|| c | c  | c ||}
  \hline
       \multicolumn{3}{||c||}{\textbf{FSO Link}}\\ \hline \hline    
 Symbol & Definition & Value \\ \hline \hline
  ${P}^{\mathrm{fso}}_{m}$ & FSO transmit power of RU$_m$ & $13$ dBm \\ \hline 
  $\delta^2$ & Noise variance at the FSO receivers & $10^{-14}$ $\mathrm{A}^2$ \\ \hline 
   $\lambda^{\mathrm{fso}}$ & Wavelength of FSO signal & $1550$ nm  \\ \hline
      $W^{\mathrm{fso}}$ & Bandwidth of FSO signal & $1$ GHz  \\ \hline   
       $R$ & Responsivity of FSO PD & $0.5\frac{1}{\mathrm{V}}$   \\ \hline  
       $\phi$ & Laser divergence angle  & $2$ mrad  \\ \hline  
       $r$ & Aperture radius  & $10$ cm  \\ \hline      
              $(\Theta,\Phi)$ & Parameters of GGamma fading  & $(2.23,1.54)$  \\ \hline\hline         
\end{tabular} 
}
\end{minipage}
\vspace{-0.8cm}
\end{table}
Unless stated otherwise, the values of the parameters for the RF and FSO links given in Table~I are used in our simulations. The noise power at the RF receivers is given by $[\varrho^2]_{\mathrm{dB}}=[\sigma^2]_{\mathrm{dB}}=W^\mathrm{rf}N_{0}+N_{F}$, where $N_{0}$ and $N_{F}$ are defined in Table~I. For simulation, we generate random fading realizations for $10^3$ fading blocks and compute the weighted sum rate of the MUs for the solution found with Algorithms~1 and 2. Moreover, we adopt $\epsilon=0.01$, and $\varepsilon=0.01$ Mbps in Algorithms~1~and~2.

\begin{figure*}[!tbp]
  \centering
  \begin{minipage}[b]{0.47\textwidth}
  \centering
\resizebox{1\linewidth}{!}{\psfragfig{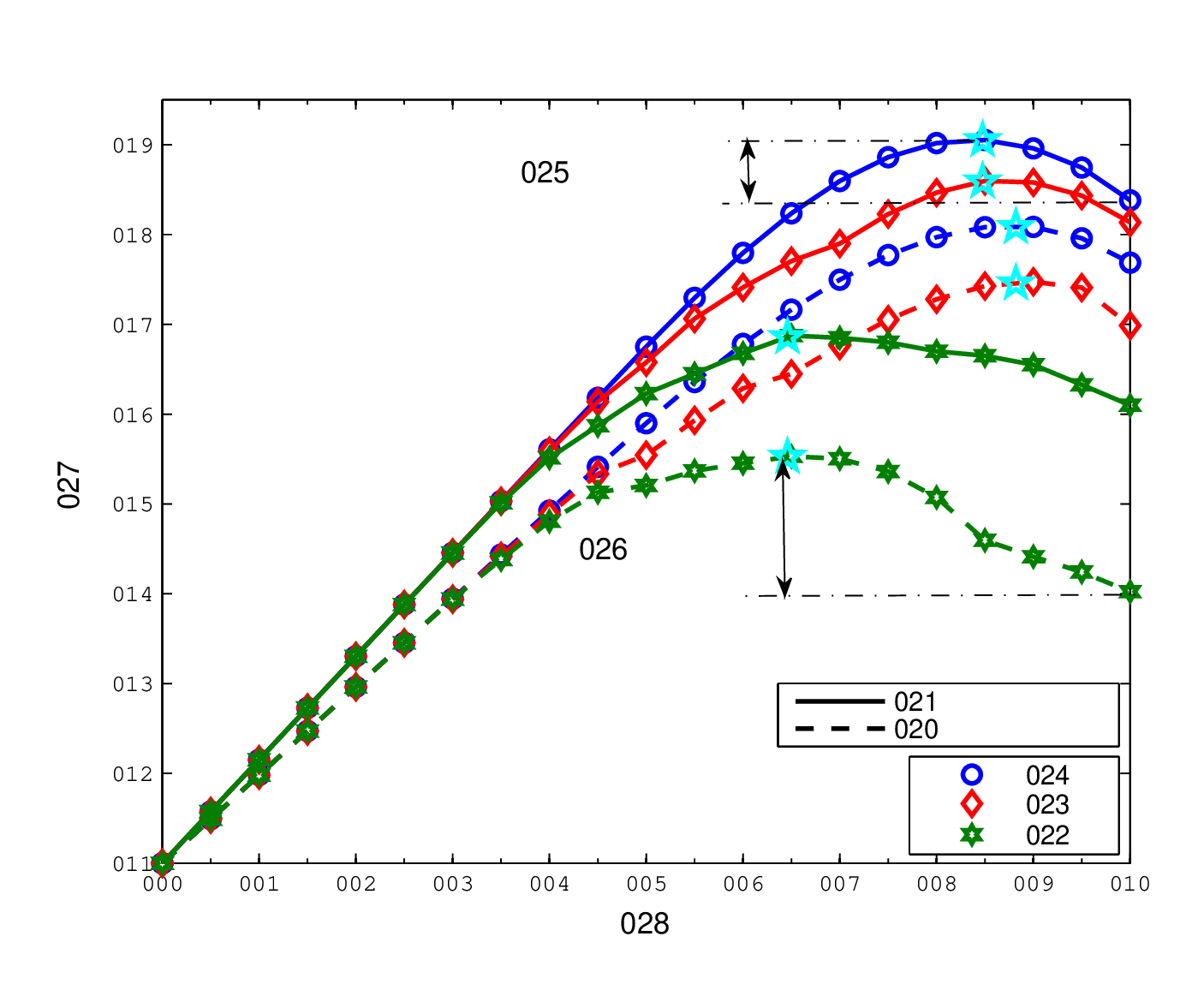} }   \vspace{-10mm}
\caption{Sum rate vs. $\alpha_0$ for $\kappa=50\times 10^{-3}, M=2, N=K=8,$ and $L=64$. Cyan star markers indicate the optimal $\alpha_0^*$ found with GSS in Algorithm~1.} 
\label{Fig:SumRate_alpha_k50}
  \end{minipage}
    \hfill
  \begin{minipage}[b]{0.1\textwidth}
  \end{minipage}
  \hfill
  \begin{minipage}[b]{0.47\textwidth}
  \centering
\resizebox{1\linewidth}{!}{\psfragfig{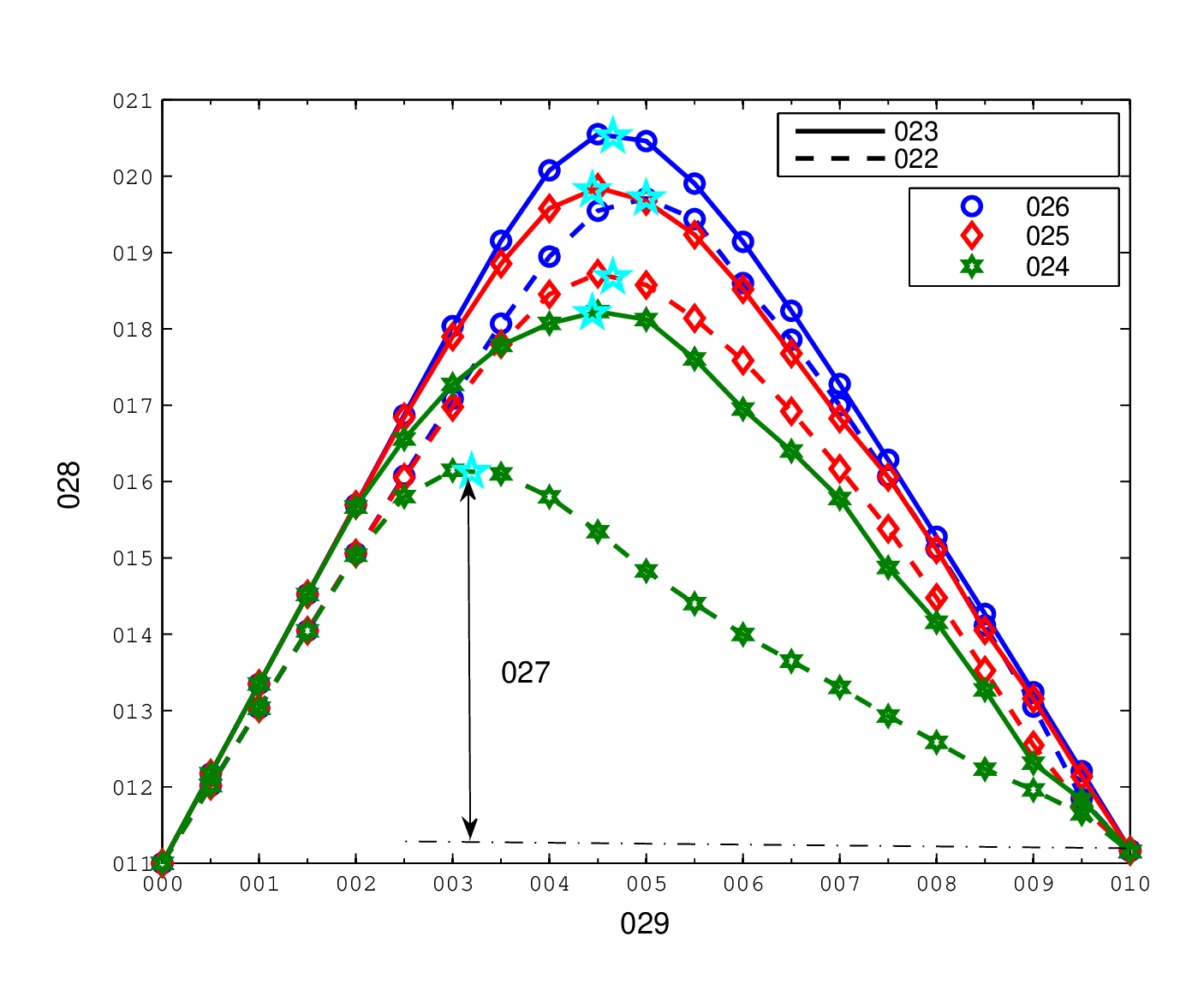} }  \vspace{-10mm}
\caption{Sum rate vs. $\alpha_0$ for $\kappa=80\times 10^{-3}, M=2, N=K=8,$ and $L=64$. Cyan star markers indicate the optimal $\alpha_0^*$ found with GSS in Algorithm~1.} 
\label{Fig:SumRate_alpha_k80}
  \end{minipage}
    \hfill
  \begin{minipage}[b]{0.02\textwidth}
  \end{minipage} \vspace{-5mm}
\end{figure*}

\subsection{Performance Evaluation}
In Figs.~\ref{Fig:SumRate_alpha_k50} and \ref{Fig:SumRate_alpha_k80}, we plot the achievable sum rate of the system vs. $\alpha_0$ for each pair of the considered quantization and detection schemes for a specific realization of the channels for two weather conditions, namely moderate and heavy fog, i.e., $\kappa=50\times10^{-3}$ and  $\kappa=80\times10^{-3}$. Figs.~\ref{Fig:SumRate_alpha_k50} and \ref{Fig:SumRate_alpha_k80} confirm the unimodality property of the sum rate w.r.t. $\alpha_0$, cf. Section~IV.B. Therefore, GSS can be employed to find the optimal fraction of the RF time slot for the access link, $\alpha_0^*$, which is denoted by cyan star markers in the figures. Figs.~\ref{Fig:SumRate_alpha_k50} and \ref{Fig:SumRate_alpha_k80} reveal that for a given detector at the CU, $\alpha_0^*$ for AVQ is smaller than that for RVQ which in turn is generally smaller than $\alpha_0^*$ for DSC. In other words, less efficient compression schemes at the RUs leads to larger output rates at the quantizers, and as a consequence, more fronthaul capacity is needed. In this case, more RF bandwidth is needed for fronthauling and less RF bandwidth is available for access. Moreover, for the adopted quantization and detection schemes, $\alpha_0^*$ is larger when the weather conditions are better, i.e., $\kappa$ is smaller, cf. Fig.~\ref{Fig:SumRate_alpha_k50}, since less RF bandwidth is needed for fronthauling as the FSO capacity is larger for better weather conditions. Furthermore, as expected, the SIC detector at the CU yields a higher achievable sum rate than the linear MMSE detector for a given RU quantization scheme. Additionally, as expected, for a given detector, DSC outperforms both RVQ and AVQ and RVQ outperforms AVQ. Finally, $\alpha_0=1$ corresponds to a system with pure FSO fronthauling. The achievable sum rate of such a system is smaller than that of the considered system with hybrid RF/FSO fronthauling.  Thereby, the gain that a hybrid RF/FSO system can achieve compared to the FSO-only system is more than $130$ and $500$~Mbps for $\kappa=50\times10^{-3}$ and  $\kappa=80\times10^{-3}$, respectively. In other words, employing hybrid RF/FSO links for fronthauling improves the performance of the system in terms of the achievable rate and ensures a non-zero minimum achievable rate even in bad weather conditions, e.g. $\kappa=80\times10^{-3}$.
%


\begin{figure*}[!tbp]
  \centering
  \begin{minipage}[b]{0.47\textwidth}
  \centering
  \resizebox{1\linewidth}{!}{\psfragfig{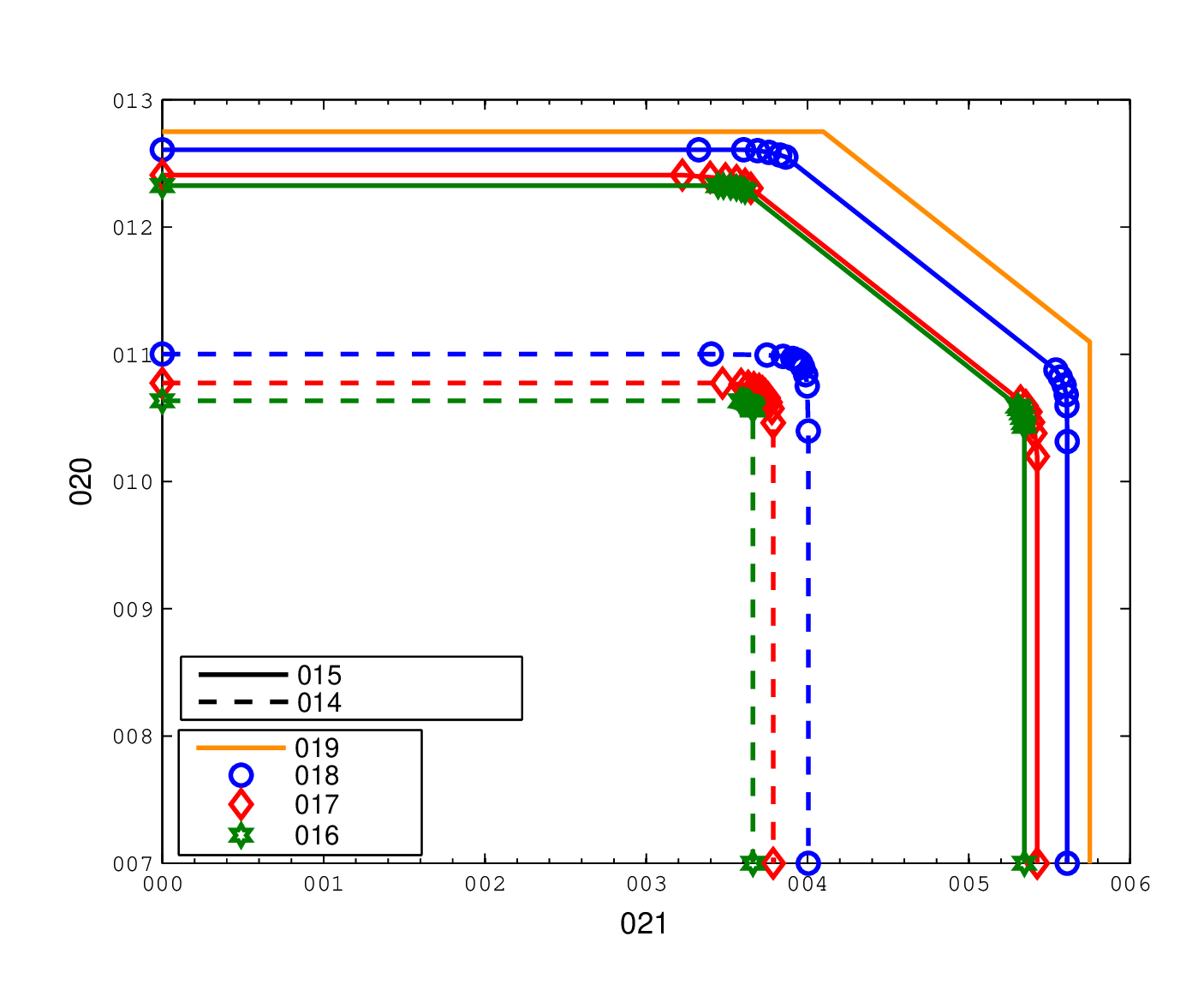} }   \vspace{-10mm}
\caption{Achievable rate region for $\kappa=50\times 10^{-3},M=N=K=L=2$.}
\label{Fig:RateRegionk50}
  \end{minipage}
    \hfill
  \begin{minipage}[b]{0.1\textwidth}
  \end{minipage}
  \hfill
  \begin{minipage}[b]{0.47\textwidth}
  \centering
\resizebox{1\linewidth}{!}{\psfragfig{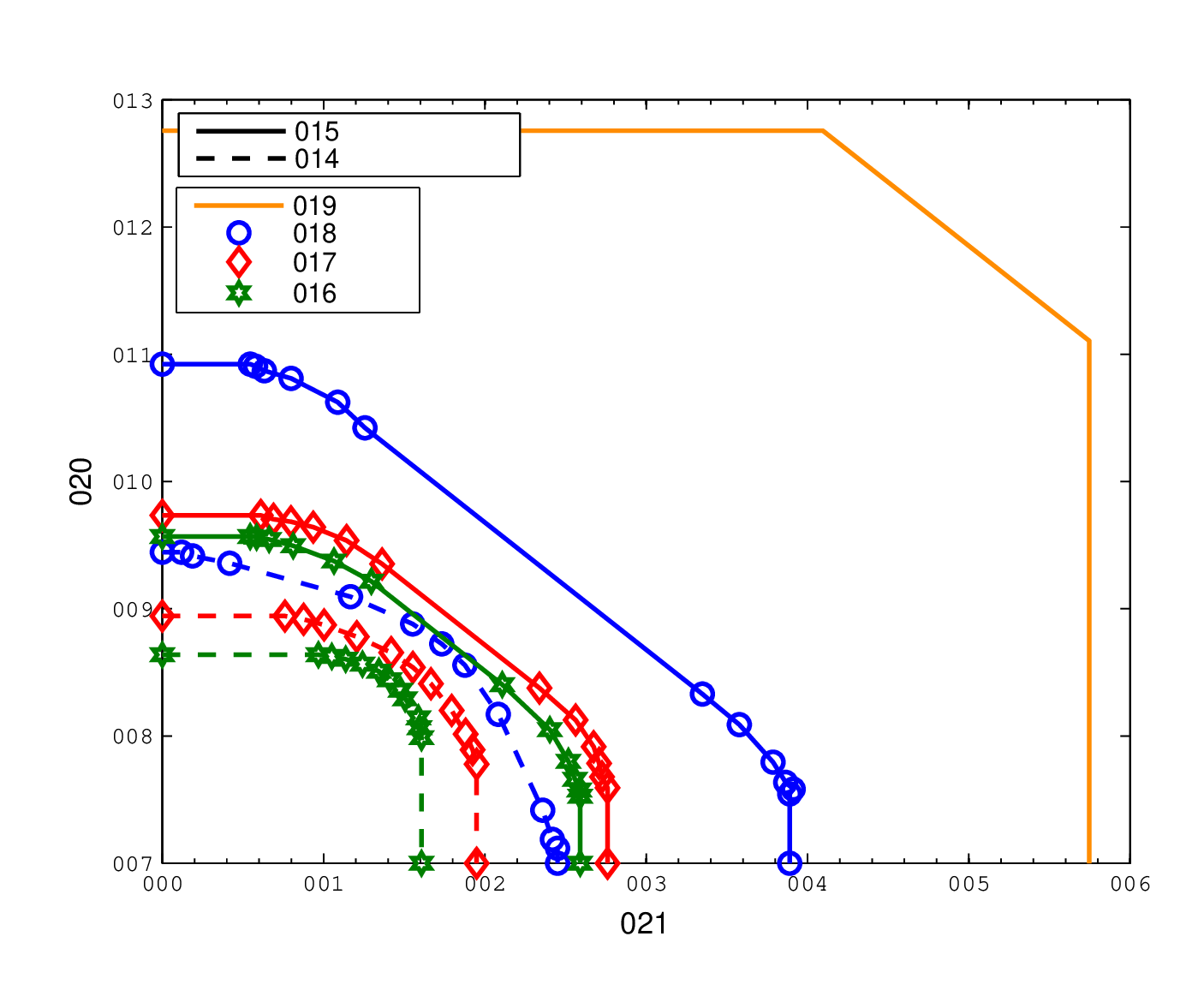} }   \vspace{-10mm}
\caption{Achievable rate region for $\kappa=80\times 10^{-3},M=N=K=L=2$.}
\label{Fig:RateRegionk80}
  \end{minipage}
    \hfill
  \begin{minipage}[b]{0.02\textwidth}
  \end{minipage} \vspace{-6mm}
\end{figure*}

Figs.~\ref{Fig:RateRegionk50} and \ref{Fig:RateRegionk80} show the achievable rate region of the considered system for  $\kappa=50\times 10^{-3}$ and $\kappa=80\times 10^{-3}$, respectively, and a non-achievable upper bound, i.e., the rate region of the virtual MAC (V-MAC) for which $\mathbf{D}=\mathbf{0}$ holds ($C_m^{\mathrm{fso}}\to\infty$). Comparing Figs.~\ref{Fig:RateRegionk50} and \ref{Fig:RateRegionk80}, we observe that when the weather conditions are favorable, cf. Fig.~\ref{Fig:RateRegionk50}, the performance difference between the considered quantization schemes is small since the FSO capacity is large, such that a more efficient compression of the signals offers diminishing performance gains. Moreover, in Fig.~\ref{Fig:RateRegionk50}, the optimal detector, i.e., SIC, approaches the rate region of the V-MAC and provides a considerable gain compared to the linear MMSE detector. On the other hand, when the weather conditions are unfavorable, cf. Fig.~\ref{Fig:RateRegionk80}, since the FSO channel capacity is small, adopting a more efficient quantization method has a significant impact on performance. This is due to the fact that utilizing a large RF time interval for fronthauling comes at the expense of reducing the RF time interval for the access links which in turn decreases the MUs' transmission rates.   
Finally, comparing Figs.~\ref{Fig:RateRegionk50} and \ref{Fig:RateRegionk80}, we observe that the better the weather conditions are, i.e., the smaller $\kappa$ is, the larger the achievable rate region becomes which is due to the increased fronthaul capacity.

\begin{figure}
  \centering
\resizebox{0.5\linewidth}{!}{\psfragfig{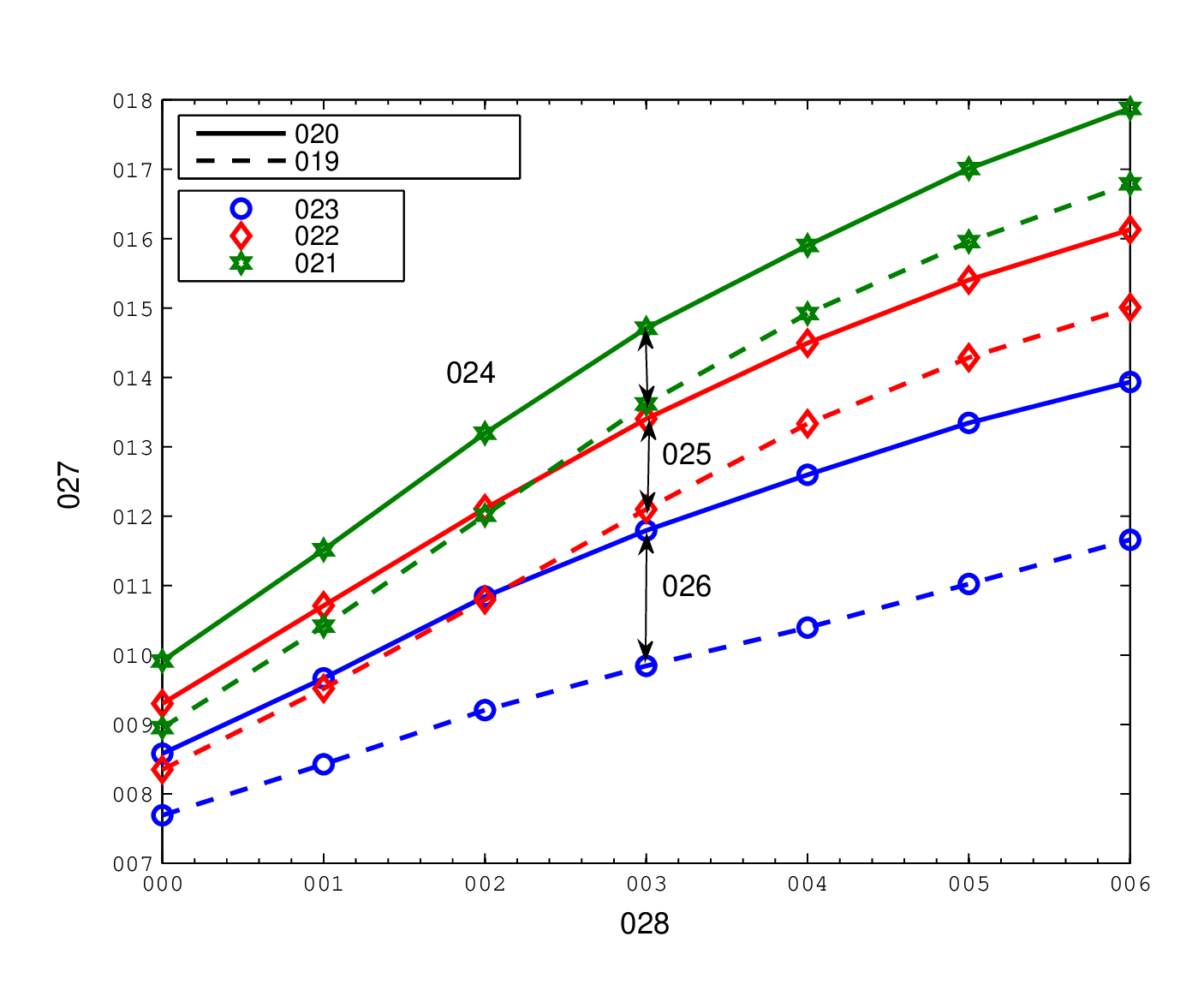} }  \vspace{-10mm} 
\caption{Average sum rate vs. MUs' transmit power $P$ for $M=2$, $K=8$, $L=64$, $\kappa=80\times 10^{-3}$, and RVQ.}
\label{Fig:SumRate_P_N}
\end{figure}

For clarity of presentation, we focus on RVQ in the following. Fig.~\ref{Fig:SumRate_P_N} shows the average sum rate vs. the transmit power of the MUs $P_1=\dots=P_M=P$ for RVQ and different numbers of RU antennas. From Fig.~\ref{Fig:SumRate_P_N}, we observe that as $P$ increases, the average sum rate increases; however, the slope of the sum rate curves decreases. The reason for this behavior is that by increasing $P$, the distortion noise also increases since the fronthaul capacity does not change. In fact, as $P\to \infty$, the fronthaul channel becomes the bottleneck and the sum rate ultimately converges to a constant value (this happens at very high SNRs outside the power range considered in Fig.~\ref{Fig:SumRate_P_N}). Moreover, we observe that the optimal SIC detector outperforms the linear MMSE detector; however, the performance gain decreases as $N$ increases, e.g. at $P=15$~dBm, SIC achieves a sum rate gain of $190$, $130$, and $110$~Mbps for $N=4$, $N=6$, and $N=8$, respectively. This is due to the fact that, as the number of receive antennas increases, linear detectors become asymptotically optimal \cite{Robert_book,MIMO}.

Next, we study how the complexity of the considered quantization and detection schemes scales in terms of the total number of antennas at the RUs, $x=MN$. Moreover, we consider the following three scenarios. For Scenario~1, we fix  $M=2$ and vary $N$, for Scenario~2, we fix $N=2$ and vary $M$, and for Scenario~3, we vary both $M$ and $N$ for $M=N$. The complexity orders of the quantization and detection schemes are given in (\ref{Eq:Complexity_Q}) and (\ref{Eq:Complexity_MMSE}), respectively, in form of $\mathcal{O}(f^{\text{q}}(M,N))$, where $f^{\text{q}}(M,N)$ is a function of $M$ and $N$ and $\text{q}\in\{\text{AVQ, RVQ, DSC}\}$ indicates the quantization scheme. Hereby, we introduce a relative complexity metric w.r.t. the baseline case of $M=N=2$ for AVQ as $f^{\text{q}}(M,N)/f^{\text{AVQ}}(2,2)$. 

\begin{figure*}[!tbp]
  \centering
  \begin{minipage}[b]{0.47\textwidth}
  \centering
\resizebox{1\linewidth}{!}{\psfragfig{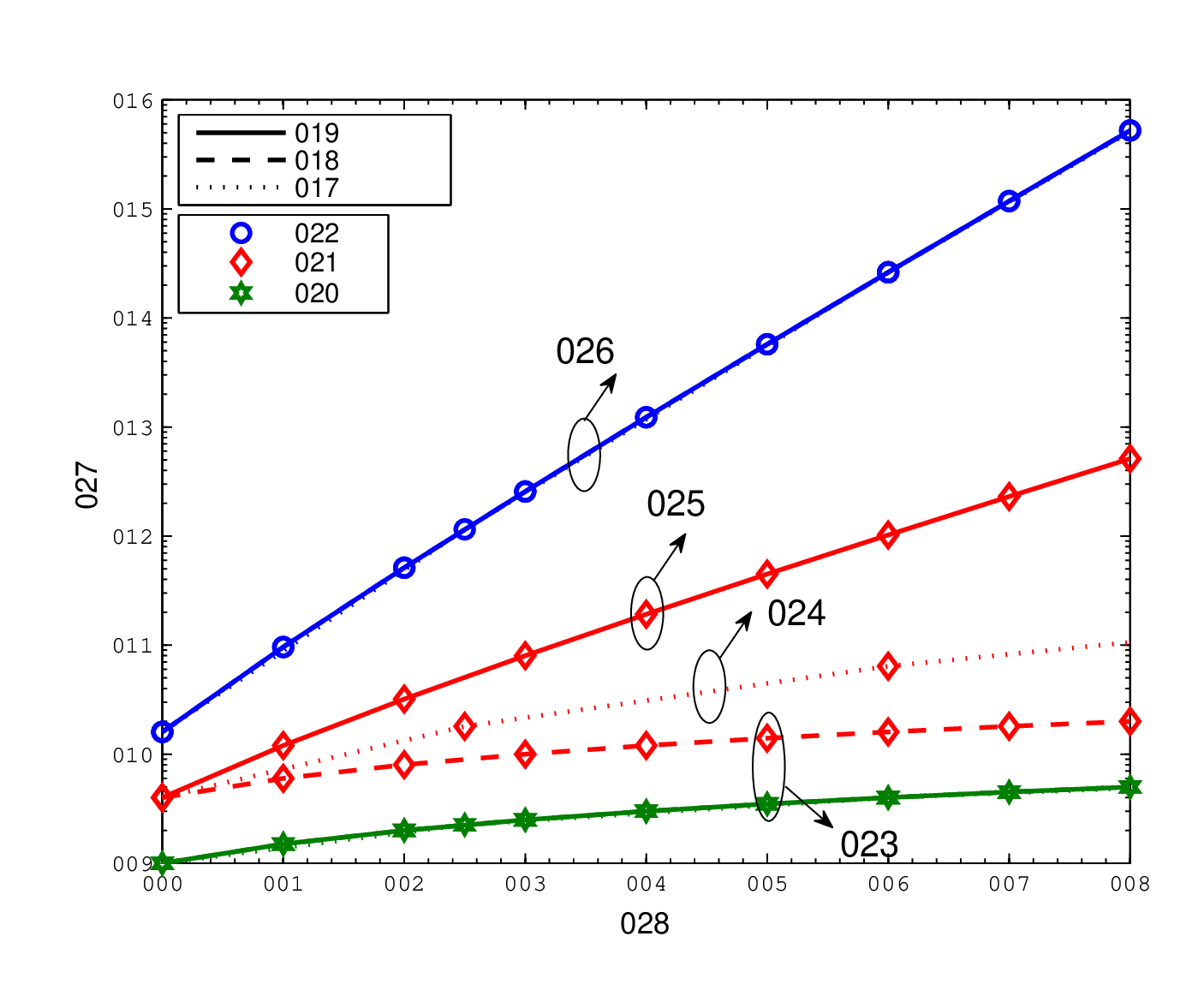} }   \vspace{-10mm}
\caption{Relative complexity of the quantization schemes vs. total number of  antennas at the RUs.}
\label{Fig:Complexity_Q}
  \end{minipage}
    \hfill
  \begin{minipage}[b]{0.1\textwidth}
  \end{minipage}
  \hfill
  \begin{minipage}[b]{0.47\textwidth}
  \centering
\resizebox{1\linewidth}{!}{\psfragfig{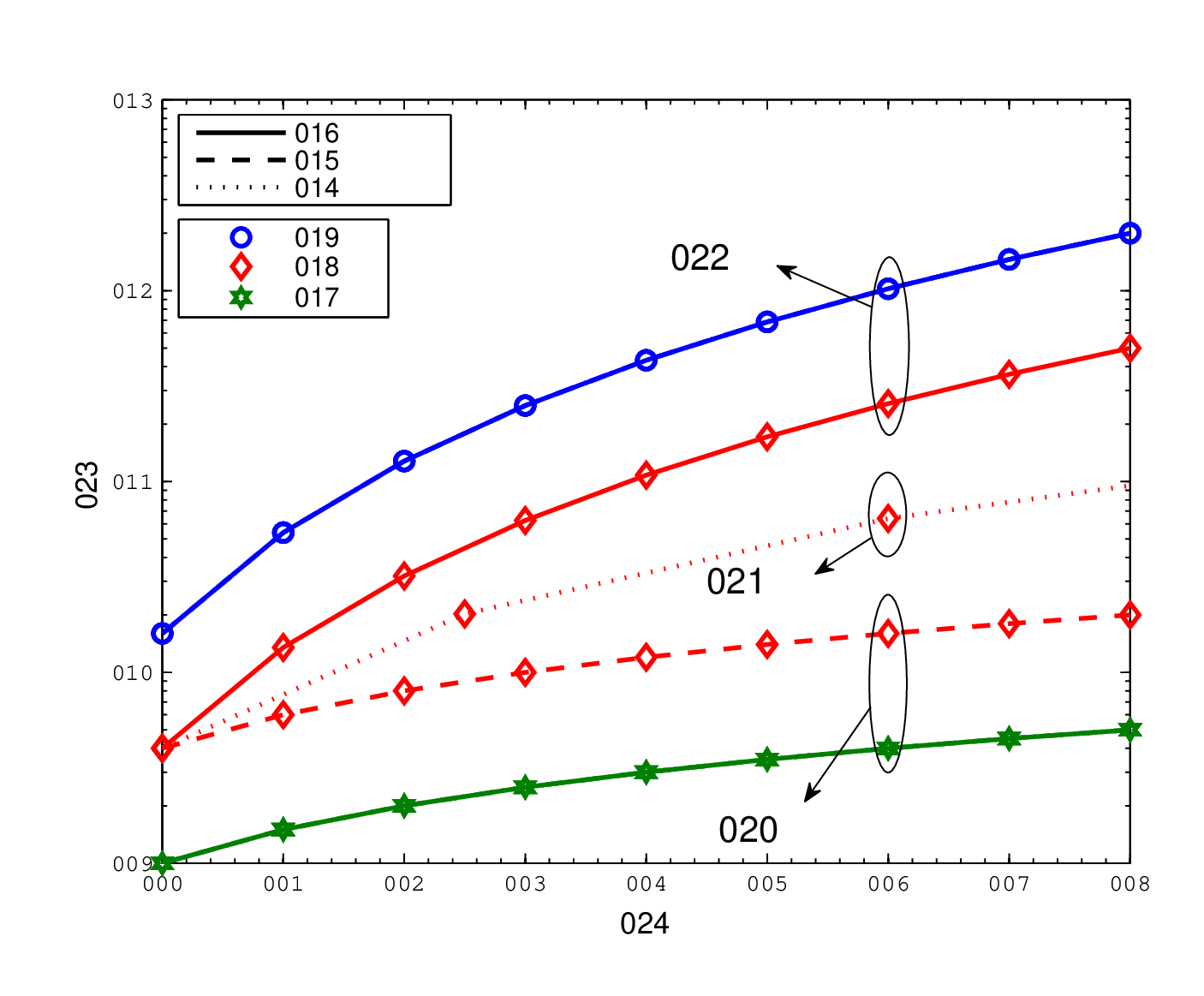} }  \vspace{-10mm} 
\caption{Relative complexity of the MMSE / SIC receiver vs. total number of antennas at the RUs for different quantization schemes.} 
\label{Fig:Complexity_D}
  \end{minipage}
    \hfill
  \begin{minipage}[b]{0.02\textwidth}
  \end{minipage} \vspace{-6mm}
\end{figure*}

Fig.~\ref{Fig:Complexity_Q} shows the relative complexity of the adopted quantization schemes vs. the total number of antennas at the RUs, $x$, for the considered three scenarios. From Fig.~\ref{Fig:Complexity_Q}, we observe that the relative complexities of AVQ and DSC are scenario independent. This is expected since for AVQ, the signal at each antenna is quantized independently, and for DSC, the signals at all antennas of the RUs are quantized jointly. On the contrary, the relative complexity of RVQ is scenario dependent. In particular, the relative complexity is higher when the number of antennas per RU is larger, i.e., the relative complexity of Scenario~1 is higher than that of Scenario~3 which is in turn higher than that of Scenario~2. Furthermore, the complexity is linear\footnote{For clarity of presentation, we plotted Figs.~\ref{Fig:Complexity_Q} and \ref{Fig:Complexity_D} in a logarithmic scale.} in $x=MN$ for AVQ for all scenarios and for RVQ for Scenario~2, whereas for all remaining cases, the complexity is exponential as shown by the big-O notation in the figure. 

Although the SIC receiver is more complex than the linear MMSE, their \textit{complexity order} in terms of the numbers of the RUs, $M$, and RU antennas, $N$, are \textit{identical}. However, their complexity order depends on the adopted quantization scheme, cf. (\ref{Eq:Complexity_MMSE}). In Fig.~\ref{Fig:Complexity_D}, we investigate the relative complexity of the considered receivers for different quantization schemes. In particular, in this figure, we show the relative complexity of the linear MMSE / SIC receiver vs. the total number of antennas at the RUs, $x$, for the considered three scenarios. As can be seen from Fig.~\ref{Fig:Complexity_D}, the relative complexity required for online detection is polynomial in $x=MN$ for all adopted quantization schemes. More specifically, the complexity is linear in $x=MN$ for AVQ for all scenarios and for RVQ for Scenario~2, quadratic for RVQ for Scenario~3, and cubic for RVQ for Scenario~1 and DSC, as shown in big-O notation in the figure.


\section{Conclusion}
In this paper, we considered an uplink C-RAN comprised of several MUs, several RUs, and one CU. The signals received at the RU antennas are compressed and forwarded to the CU over hybrid RF/FSO fronthaul links. Since the RF resources are limited, we assumed that the  multiple access and fronthaul links share the same RF resources in an orthogonal manner. We further considered three quantization schemes, namely AVQ, RVQ, and DSC at the RUs and two detection schemes, namely linear MMSE and SIC at the CU in order to strike a trade-off between performance and complexity. We proposed an algorithm to efficiently jointly optimize the quantization noise covariance matrices at the RUs and the RF time interval size of the RF multiple access and fronthaul links for maximization of the rate region. To arrive at this algorithm, we first formulated a unified weighted sum rate maximization problem valid for each pair of the considered quantization and detection schemes. Then, to overcome the non-convexity of the original problem, we transformed it into a bi-convex problem which enabled the design of the proposed algorithm for finding an efficient suboptimal solution. Moreover, we analyzed the asymptotic complexities of the proposed algorithm and the adopted quantization and detection schemes. Our simulation results showed that, for adverse weather conditions, DSC provides a considerable performance gain over RVQ and similarly RVQ over AVQ. In contrast, for favorable weather conditions, the performance difference between these quantization schemes becomes small, and the system performance for SIC approaches that of the V-MAC upper bound. Furthermore, the performance difference between the SIC and linear MMSE receivers becomes small for large numbers of RU antennas. Finally, our simulation results revealed that hybrid RF/FSO fronthauling outperforms FSO-only fronthauling, especially when the FSO links experience adverse atmospheric conditions.\vspace{-0.4cm}


\appendices
\section{}\label{App:Pro}
In the following,  we rewrite the objective function, source coding constraint $\mathrm{C}1$, and the constraint for distortion matrix $\mathbf{D}$ in (\ref{Eq:RateRegion}) in a unified manner for all considered quantization and detection schemes. Note that the channel coding constraint $\mathrm{C}2$ in (\ref{Eq:RateRegion}) does not depend on the quantization and detection schemes.

\textit{Objective function:} The objective function of (\ref{Eq:RateRegion}) depends on the adopted detection scheme, namely linear MMSE and SIC, and can be computed based on (\ref{Eq:R_k_MMSE}) and (\ref{Eq:R_k_S}), respectively. Defining constant matrices $\mathbf{V}_k$ in (\ref{Eq:Sets}) and $\mathbf{W}_k$ in (\ref{Eq:Sets}) for the MMSE and SIC receivers, the objective function can be written in the unified form given in (\ref{Eq:Unified}).
 
\textit{Constraint $\mathrm{C}1$:} The source coding constraints for AVQ, RVQ, and DSC are given in (\ref{Eq:SQ_SC}), (\ref{Eq:VQ_SC}), and (\ref{Eq:DSC_SC}), respectively. Note that each of these constraints affect only cetrain elements of distortion matrix $\mathbf{D}$. To extract the required submatrices from $\mathbf{D}$, we employ the linear operation $\mathbf{I}(\mathcal{T}_{\mathcal{S}})\mathbf{D}\mathbf{I}(\mathcal{T}_{\mathcal{S}})^{\mathsf{T}},\,\forall\mathcal{S}\in\bar{\mathcal{S}}$, which extracts the elements of $\mathbf{D}$ whose row and column indices belong to $\mathcal{T}_{\mathcal{S}}$. Thereby, by properly defining  $\bar{\mathcal{S}}$ for the considered quantization schemes as in (\ref{Eq:Sets}), the source coding constraint can be written in a unified manner, see (\ref{Eq:Unified}). 

\textit{Distortion matrix:} Note that different quantization schemes enforce different constraints on $\mathbf{D}$, namely $\mathbf{D}$ is diagonal, block diagonal, and a general covariance matrix for AVQ, RVQ, and DSC, respectively. Therefore, we impose these properties by employing unified constraints $\mathbf{I}(\mathcal{T})\mathbf{D}\mathbf{I}(\mathcal{T})^{\mathsf{T}}\succeq\mathbf{0}$ and $\mathbf{I}(\mathcal{T})\mathbf{D}\mathbf{I}(\mathcal{T}^{\mathsf{c}})^{\mathsf{T}}=\mathbf{0}_{|\mathcal{T}|\times|\mathcal{T}^{\mathsf{c}}|},\,\forall\mathcal{T}\in\bar{\mathcal{T}}$ and by properly defining $\bar{\mathcal{T}}$ for the considered quantization schemes in (\ref{Eq:Sets}). This completes the proof.

\section{}\label{App:Lem}
Dividing both sides of constraint $\mathrm{C}2$ in (\ref{Eq:Unified}) by $C_m^{\mathrm{rf}}$ and summing the right-hand sides and left-hand sides of all constraints over indices $m\in\mathcal{S}$,  where $\mathcal{S}$ is a non-empty subset of $\mathcal{M}$, we obtain
\begin{IEEEeqnarray}{lll}\label{Eq:a}
\sum_{m\in\mathcal{S}}\frac{r_m}{C_m^{\mathrm{rf}}} \leq \sum_{m\in\mathcal{S}} \alpha_m + \sum_{m\in\mathcal{S}}\frac{C_m^{\mathrm{fso}}}{C_m^{\mathrm{rf}}}.
\end{IEEEeqnarray}
After some manipulations,  we obtain
\begin{IEEEeqnarray}{lll}\label{Eq:b}
 \frac{ \sum_{m\in\mathcal{S}}\Big(r_m\prod_{m'\neq m, m'\in\mathcal{S}}C_{m'}^{\mathrm{rf}}\Big)}{ \prod_{m\in\mathcal{S}}C_m^{\mathrm{rf}}}
\leq \sum_{m\in\mathcal{S}} \alpha_m + \frac{ \sum_{m\in\mathcal{S}}\Big(C_m^{\mathrm{fso}}\prod_{m'\neq m, m'\in\mathcal{S}}C_{m'}^{\mathrm{rf}}\Big)}{\prod_{m\in\mathcal{S}}C_{m}^{\mathrm{rf}}},
\end{IEEEeqnarray}
which can be rewritten as follows
\begin{IEEEeqnarray}{lll}\label{Eq:c}
 \sum_{m\in\mathcal{S}}r_m G_m(\mathcal{S}) \leq (1-\alpha_0)G(\mathcal{S}) + \sum_{m\in\mathcal{S}}C_m^{\mathrm{fso}} G_m(\mathcal{S}),\quad
\end{IEEEeqnarray}
where $G_m(\mathcal{S})=\frac{\prod_{\forall m' \in \mathcal{S}}C_{m'}^{\mathrm{rf}}}{C_m^{\mathrm{rf}}}$ , $G(\mathcal{S})=\prod_{\forall m\in \mathcal{S}}{C}_m^{\mathrm{rf}}$. Moreover, we use the inequality $\sum_{m\in\mathcal{S}} \alpha_m\leq 1-\alpha_0$  which in general enlarges the corresponding  feasible set compared to that for the original constraints. However, the original feasible set defined by the constraint in (\ref{Eq:Unified}) for $\forall m\in\mathcal{M}$ is  identical to the feasible set of inequality (\ref{Eq:c}) if all $\mathcal{S}\subseteq \mathcal{M}$ are considered~\cite[Chapter~15]{Cover}. This completes the proof.

\section{}\label{App:Unim}
In the following, we first establish why it is in general challenging to mathematically show the unimodality of the objective function in (\ref{Eq:Unified}) w.r.t. $\alpha_0$. Note that the objective function in (\ref{Eq:Unified}) is a function of both $\alpha_0$ and $\mathbf{D}$. To show that the objective function is a unimodal function of $\alpha_0$ for the optimal $\mathbf{D}^*$, we would have to first establish an analytical relation between $\alpha_0$ and $\mathbf{D}^*$. However, such an analytical relation does not exist in general. Therefore, we focus on the special case of $M=N=K=1$ where the scalar distortion $d^*$ can be uniquely found as a function of $\alpha_0$. In particular, in this case, (\ref{Eq:Unified}) simplifies~to\footnote{For notational simplicity, we drop here the user index, $k$, the RU index, $m$, and the antenna index, $n$. Moreover, since we only have one RU and one antenna, the distortion noise is a scalar resulting in a scalar distortion noise variance $d$.} 
\begin{IEEEeqnarray}{lll}\label{Eq:R}
&\underset{\alpha_0\in [0,1], d\geq 0}{\mathrm{maximize}}\,\,& R(\alpha_0,d)\triangleq\alpha_0W^{\mathrm{rf}}\log_2\left(\frac{P|h|^2+d+\sigma^2}{d+\sigma^2}\right) \\
 &\mathrm{subject\,\, to\,\,} &F(\alpha_0,d)\triangleq\alpha_0f_s\log_2\left(\frac{P|h|^2+d+\sigma^2}{d}\right)-(1-\alpha_0)C^{\mathrm{rf}}-C^{\mathrm{fso}}\leq0.\nonumber
\end{IEEEeqnarray}
For a given $\alpha_0$, the optimal $d^*$ can be shown to satisfy the constraint with equality, i.e., $F(\alpha_0,d^*)=0$. Let us define $g(d^*)=W^{\mathrm{rf}}\log_2\left(\frac{P|h|^2+d^*+\sigma^2}{d^*+\sigma^2}\right)$ and $f(d^*)=f_s\log_2\left(\frac{P|h|^2+d^*+\sigma^2}{d^*}\right)$. To show the unimodality of $R(\alpha_0,d^*)$, we have to show that $\frac{\partial R(\alpha_0,d^*)}{\partial \alpha_0}$ changes its sign at most once for $\alpha_0\in[0,1]$.  Using the chain rule, the derivative of $R(\alpha_0,d^*)$ w.r.t. $\alpha_0$ is obtained~as
\begin{IEEEeqnarray}{lll}\label{Eq:R_derivative}
\frac{\partial R(\alpha_0,d^*)}{\partial \alpha_0}=g(d^*)+\alpha_0\frac{\partial g(d^*)}{\partial d^*}\times\frac{\partial d^*}{\partial \alpha_0},
\end{IEEEeqnarray}
where $\frac{\partial g(d^*)}{\partial d^*}=-\frac{W^{\mathrm{rf}}P|h|^2}{\ln(2)(d^*+\sigma^2)(P|h|^2+d^*+\sigma^2)}$. Moreover, $\frac{\partial d^*}{\partial \alpha_0}$ is obtained by taking the derivative of $F(\alpha_0,d^*)=0$, which yields $\frac{\partial d^*}{\partial \alpha_0}=-\frac{C^{\mathrm{rf}}+f(d^*)}{\alpha_0}\left(\frac{\partial f(d^*)}{\partial d^*}\right)^{-1}$, where  $\frac{\partial f(d^*)}{\partial d^*}=-\frac{f_s(P|h|^2+\sigma^2)}{\ln(2)d^*(P|h|^2+d^*+\sigma^2)}$. Substituting these results in (\ref{Eq:R_derivative}), we obtain $\frac{\partial R(\alpha_0,d^*)}{\partial \alpha_0}$ as follows
\begin{IEEEeqnarray}{lll}\label{Eq:R_derivative_Final}
\frac{\partial R(\alpha_0,d^*)}{\partial \alpha_0}=g(d^*)-c\left(f(d^*)+C^{\mathrm{rf}}\right)\frac{d^*}{d^*+\sigma^2}.
\end{IEEEeqnarray}
where $c=\frac{W^{\mathrm{rf}}P|h|^2}{f_s\left(P|h|^2+\sigma^2\right)}$ is a positive constant. Note that $\frac{\partial R(\alpha_0,d^*)}{\partial \alpha_0}|_{\alpha_0=0}=W^{\mathrm{rf}}\log_2\left(\frac{P|h|^2+\sigma^2}{\sigma^2}\right)\geq0$ where we exploited the fact that $d^*=0$ holds for $\alpha_0=0$. Therefore, it suffices to show that $\frac{\partial R(\alpha_0,d^*)}{\partial \alpha_0}$ is monotonically decreasing, i.e., $\frac{\partial^2 R(\alpha_0,d^*)}{\partial \alpha_0^2}\leq0$, to ensure that $\frac{\partial R(\alpha_0,d^*)}{\partial \alpha_0}$ changes its sign at most once. In particular, by taking the derivative of (\ref{Eq:R_derivative_Final}) w.r.t. $\alpha_0$, we obtain
\begin{IEEEeqnarray}{lll}\label{Eq:R_Second_derivative}
\frac{\partial^2 R(\alpha_0,d^*)}{\partial \alpha_0^2}=-c\left(f(d^*)+C^{\mathrm{rf}}\right) \frac{\sigma^2}{(d^*+\sigma^2)^2}\frac{C^{\mathrm{rf}}+f(d^*)}{\alpha_0}\frac{\ln(2)d^*(P|h|^2+d^*+\sigma^2)}{f_s(P|h|^2+\sigma^2)}\leq0.
\end{IEEEeqnarray}
 This implies that for the considered special case,  $R(\alpha_0,d^*)$ is even a convex function of $\alpha_0$ and consequently also unimodal. This completes the proof.
\bibliographystyle{IEEEtran}
\bibliography{My_Citation_13-08-2018}


\usepackage [active,tightpage]{preview}
\let \document \origdocument 
\pagestyle {empty}
\begin {document}
\centering \null \vfill 
\begin {preview}
 \psfrag {Csum}[c][c][1]{$T$} \psfrag {al}[c][c][1]{$\alpha _0$} \psfrag {amin}[c][c][1]{$\alpha _0^{\mathrm {min}}$} \psfrag {amax}[c][c][1]{$\alpha _0^{\mathrm {max}}$} \psfrag {a1}[c][c][1]{$\alpha _0^{(1)}$} \psfrag {a2}[c][c][1]{$\alpha _0^{(2)}$} \psfrag {da}[c][c][1]{$\rho \Delta \alpha $} \psfrag {NewS}[c][c][1]{New search interval} \psfrag {C1}[c][c][1]{$T^{(1)}$} \psfrag {C2}[c][c][1]{$T^{(2)}$} 
\includegraphics [width=1.2\linewidth ] {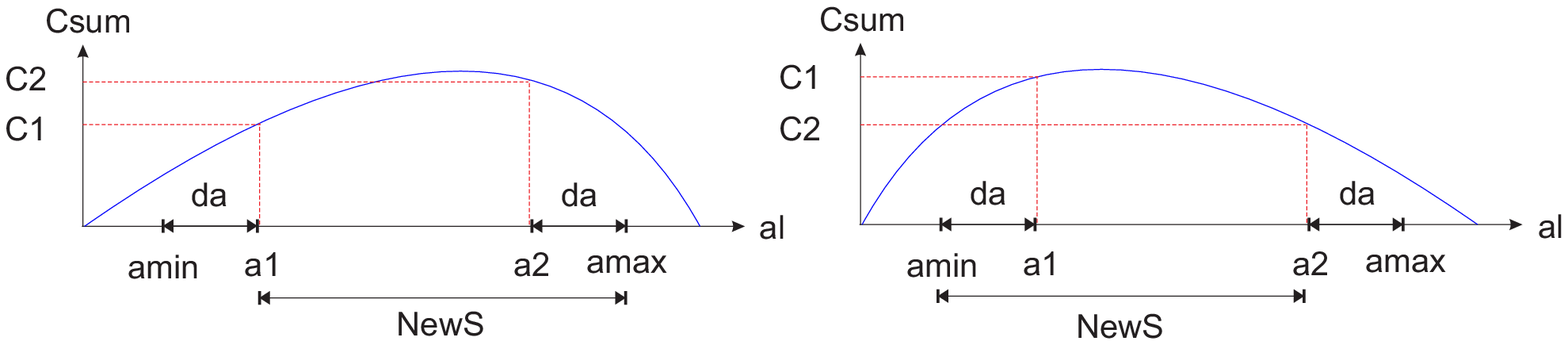}
\end {preview}
\vfill \end {document}